\documentclass[aip,jcp,amsmath,amssymb,preprint,groupedaddress]{revtex4-1}

\usepackage{graphicx}
\usepackage{dcolumn}
\usepackage{bm}

\usepackage[utf8]{inputenc}
\usepackage[T1]{fontenc}
\usepackage{mathptmx}

\usepackage{amsthm}
\usepackage{soul}
\usepackage[dvipsnames]{xcolor}
\setul{}{2pt}
\setstcolor{red}
\DeclareGraphicsExtensions{{.eps,.pdf,.png}}

\begin{document}

\preprint{JCP}

\title{Full molecular dynamics simulations of molecular liquids for single-beam spectrally controlled  two-dimensional Raman spectroscopy}

\author{Ju-Yeon Jo}
 \email{jo.juyeon.46v@kyoto-u.jp}
\author{Yoshitaka Tanimura}
 \email{tanimura.yoshitaka.5w@kyoto-u.jp}
 \affiliation{Department of Chemistry, Graduate School of Science, Kyoto University, Kyoto 606-8502, Japan}

\date{\today}

\begin{abstract}
Single-beam spectrally controlled (SBSC) two-dimensional (2D) Raman spectroscopy is a unique 2D vibrational measurement technique utilizing trains of short pulses that are generated from a single broadband pulse by pulse shaping. This approach overcomes the difficulty of 2D Raman spectroscopy in dealing with small-signal extraction and avoids complicated low-order cascading effects, thus providing a new possibility for measuring the intramolecular and intermolecular modes of molecular liquids using  fifth-order 2D Raman spectroscopy. Recently, for quantitatively investigating the mode--mode coupling mechanism, Hurwitz \emph{et al.} [Optics Express \textbf{28}, 3803 (2020)] have developed a new pulse design for this measurement  to separate the contributions of the fifth- and  third-order polarizations, which  are often overlapped in the original single-beam measurements. Here, we describe a method for simulating these original measurements and the new 2D Raman measurements on the basis of a second-order response function approach. We carry out full molecular dynamics simulations for carbon tetrachloride and liquid water using an equilibrium--nonequilibrium hybrid algorithm, with the aim of  explaining the key features of the SBSC 2D Raman spectroscopic method from a theoretical point of view. The predicted signal profiles and intensities provide valuable information that can be applied to 2D spectroscopy experiments, allowing them to be carried out more efficiently.
\end{abstract}

\maketitle

\section{\label{sec:level1}Introduction}

Vibrational motion in liquids can be categorized into intermolecular and intramolecular dynamics, which correspond to the low-frequency  and  high-frequency regimes, respectively.\cite{Mukamel}  Intramolecular dynamics are particularly important for understanding the characteristic features of dilute molecular systems, and various multidimensional vibrational spectroscopic methods have been developed for this purpose,  such as two-dimensional (2D) infrared (IR) spectroscopy.\cite{Hamm, Cho} However, these methods are not suitable for the investigation of  intermolecular dynamics, because of their limited ability to capture low-frequency vibrational modes: there are very few methods that can probe frequencies below 1000~cm$^{-1}$, where  intermolecular dynamics usually take place. 

Fifth-order 2D Raman spectroscopy is the oldest of all the multidimensional laser spectroscopic techniques and is probably one of the most difficult to implement experimentally.\cite{Inhomo01} It uses two pairs of Raman excitation pulses together with a probe pulse for generating polarizability. The time delay between the excitation pulse pairs is $t_1$, with the probe pulse following the second pair after a time $t_2$. The method is thus characterized by the two time variables $(t_1, t_2)$. It has created the possibility for quantitatively investigating the role of intermolecular vibrational motion of liquid molecules, while 2D IR spectroscopy was developed for the study of intramolecular vibrational modes of molecules in liquids and in molecules of biological significance.\cite{Hamm,Cho} Although 2D Raman spectroscopy was originally proposed to investigate the difference between homogeneous and inhomogeneous broadening,\cite{Inhomo01} it has also been successfully employed to investigate  anharmonicity of  potentials,\cite{AH01,AH02,AH03,AH04,AH05} 
mode--mode coupling mechanisms,\cite{couple01,couple02,couple03,couple04,couple05} and  dephasing processes\cite{dephase01,dephase02,dephase03,dephase04,dephase05,dephase06,dephase07,dephase08,dephase09} of   intermolecular modes. It has been recognized that in this high-order nonlinear experiment, it is extremely difficult to deal with small-signal extraction, owing to the unforeseen low-order cascading effect of light emission,\cite{cascade01,cascade02,cascade03} and such  
experimental signals have only been obtained for carbon disulfide,\cite{CS2-01,CS2-02,CS2-03,CS2-04} benzene,~\cite{C6H6-01} and formamide~\cite{HCONH2-01} liquids.  Although the 2D Raman signals of liquid water have not yet been observed, 2D THz-Raman (also known as 2D Raman-THz) spectroscopy has been investigated both theoretically and experimentally\cite{2DTR-perspective} with the aim of measuring  signals for water,\cite{2DTR-MD01,2DTR-MD02,2DTR-MD03,2DTR-EX01,2DTR-MD04,2DTR-MD05,2DTR-TM01} solvated ions,\cite{2DTR-EXA01,2DTR-EX01,2DTR-EXC01}
and carbon tetrachloride.\cite{2DTR-EXD01} In this method,  cascading effects are suppressed using two terahertz pulses and one set of Raman pulses. As an extension of 2D THz-Raman spectroscopy, 2D IR-Raman spectroscopy has also been proposed to study  interactions among intermolecular and intramolecular modes.\cite{2DIR,2DIR-EX1} Because intermolecular vibrational modes are usually both Raman- and IR-active, the information that  can be obtained from  2D Raman and 2D THz-Raman signals can be used to investigate the fundamental nature of intermolecular interactions in a complementary manner.  Nevertheless, it is desirable to develop  2D Raman techniques in which  cascading effects are suppressed, because  Raman measurements can be applied to a wide variety of materials, including glasses and solids at high resolution,\cite{Nagata2DR1,Nagata2DR2} in the range from low-frequency intermolecular modes to high-frequency intramolecular modes.

In 2015, a single-beam spectrally controlled (SBSC) 2D Raman spectroscopic method, whose signals are generated from a coherently controlled pulse, was developed to overcome the cascading problem.\cite{Frostig}  SBSC 2D Raman measurements are useful because they can be
applied to a wide variety of materials with a simple optical setup suitable for applications in ultrafast nonlinear optical 
microscopy.\cite{Frostig_signalrecovery,Silberberg1,tightfo} In such a measurement, a vibrational excitation is created by a sequence of pulses whose time period corresponds to a vibrational mode.\cite{Silberberg1} Thus, by sweeping the pulse periods, we can tune the excitation frequency of the vibrational modes (see  Appendix~\ref{s:Ramanprocess1}). The use of spectral broadband pulses provides all five pulses for the pump and probe frequencies that are necessary for  2D Raman measurements. The signal is then detected using the phase shift between the input and output electric fields arising from the coherent Stokes and anti-Stokes processes. Utilizing heterodyne detection, we can remove the unwanted cascading contributions by measuring the real part of the signal. This is possible because the cascading signal does not exhibit any frequency shift resulting from the difference in the underlying excitation mechanisms.  Although the original pulse design for such measurements had  difficulty in separating the contributions of the fifth- and  third-order polarizations, a recently introduced pulse design overcomes this problem and opens up a new possibility for SBSC 2D Raman spectroscopy.\cite{Hurwitz} Thus, we are now able to investigate complex dynamics of molecular liquids, where both intermolecular and intramolecular modes and the interaction among these modes play significant roles.

 In these measurements, however, the analysis of the spectra is more complicated than in conventional 2D Raman spectroscopy, because the polarization is  measured indirectly by using the phase shifts of the applied electric fields. Therefore, in this paper, we carry out full molecular dynamics (MD) simulations for liquid carbon tetrachloride and liquid water, with the aim of understanding the key features of this spectroscopic method from a theoretical point of view. Thereby, we hope to provide valuable information that can be applied to 2D spectroscopic  experiments, allowing them to be carried out more efficiently. In Sec.~\ref{secII}, we formulate the single-beam 1D and 2D Raman spectra on the basis of the response function approach.
In Sec.~\ref{secIII}, we explain the methodology for simulating 1D and 2D Raman signals with full MD simulations. In Sec.~\ref{secIV}, we investigate the calculated 1D and 2D Raman signals for liquid  carbon tetrachloride and liquid water to elucidate the characteristic features of intermolecular and intramolecular vibrational modes and their couplings. Section~\ref{secV} is devoted to concluding remarks.

\section{Observables of the SBSC 1D and 2D Raman measurements}\label{secII}

In standard 1D Raman measurements in the time domain, a pair of off-resonant pump pulses $E_1(t)E_1^*(t)$ are used, where
$E_1^*(t)$ is the complex conjugate of $E_1(t)$, together with a subsequent probe pulse $E_f(t)$. The resulting signal is then detected as a function of the delay time between the pump and the probe. In the 2D Raman case, the signal is generated using two pairs of off-resonant pulses $E_1(t)$ and $ E_2(t)$, followed by the probe pulse $E_f(t)$. In this case, the signal is measured as a function of two time variables, corresponding to the delay times between the two pairs of pump pulses and the probe pulse. The observables of the optical measurements in 1D and 2D Raman experiments are then defined as the third- and fifth-order polarization functions, respectively. In the classical limit, these are expressed as~\cite{Hybrid02}
\begin{align}
P^{(3)}(t) ={}& E_f(t)\int_{0}^{\infty}{dt_1}\,R^{(3)}(t_1)E_1(t-t_1)E_1^*(t-t_1), \label{e:3rd_ttime}\\[6pt]
P^{(5)}(t) = {}&E_f(t)\int_{0}^{\infty}{dt_2}\int_{0}^{\infty}{dt_1}\,R^{(5)}(t_1,t_2)E_2(t-t_2)E_2^*(t-t_2)\nonumber\\
&\times E_1(t-t_2-t_1)E_1^*(t-t_2-t_1).  \label{e:5th_ttime} 
\end{align}
Here, the third- and fifth-order response functions are defined as
\begin{align}
R^{(3)}(t_1) &\equiv \langle \{ {\Pi} (t_1), {\Pi} (0) \} \rangle,  \label{e:3rd_R}\\[6pt]
R^{(5)}(t_1,t_2) &\equiv \langle \{ \{ {\Pi}(t_1+t_2), {\Pi} (t_1)\}, \Pi (0) \} \rangle, \label{e:5th_R} 
\end{align}  
where ${\Pi}(t)$ is the molecular polarizability and $\{~,~\}$ is the Poisson bracket.  

In contrast to the signals used in conventional 1D and 2D Raman spectroscopy, which are created using ultrashort Raman pulses, those of SBSC Raman spectroscopy are, in principle, frequency-domain measurements, although in a real-time description of these measurements, they can be regarded as being composed of short pulse trains.\cite{Frostig, Frostig_signalrecovery,Silberberg1,tightfo,Hurwitz}  These measurements rely on the vibrational excitations that are created by a pulse sequence whose time period corresponds to a vibrational mode. 

\begin{figure*}[t!]
\includegraphics[width=0.5\textwidth]{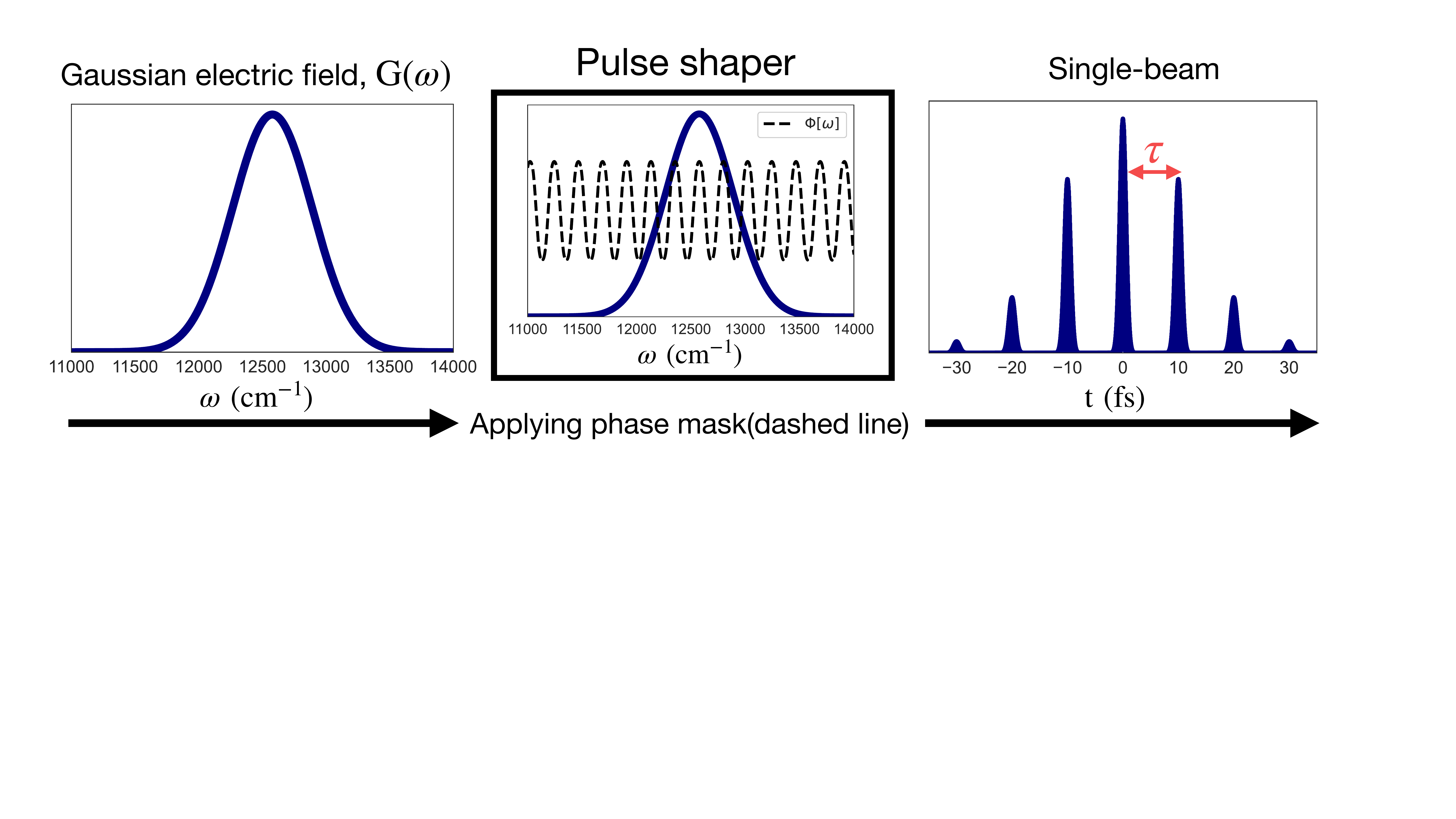}
\caption{\label{f:pshaper}Schematic illustration of the pulse shaping technique, showing the phase modulation $\phi[\omega]$ applied to the broadband Gaussian electric field pulse. }
\end{figure*}

The third-order polarizability functions are expressed in Fourier convolution form as (see Appendix~\ref{s:Ramanprocess1})
\begin{align}
P^{(3)}[\omega] 
= {}&\frac{1}{2\pi}\int_{-\infty}^{\infty}{{d\omega_1}\,{R}^{(3)}[\omega_1]{E}[\omega-\omega_1]}\nonumber\\
&\times\int_{-\infty}^{\infty}{{d\Omega}\,{E}[\Omega]{E}^*[\Omega-\omega_1]} ,  \label{e:3rd_omega}
\end{align}
where $f[\omega] = \mathcal{F}{[f(t)]}$ is the Fourier representation of any function $f(t)$, and $\Omega$  and $\Omega-\omega_1$ correspond to the pump and Stokes frequencies, respectively.
This measurement utilizes a broadband Gaussian electric field $G(\omega)$ with a modulated phase mask $\phi[\omega]$ described as\cite{Frostig}  (see Fig.~\ref{f:pshaper})
\begin{align}
{E}[\omega] = G(\omega)  \exp(i\phi^{(3)}[\omega; \tau]), 
\label{e:EwO} 
\end{align}
where 
\begin{align}
G(\omega) = E_0\left\{\exp\!\left[-\frac{(\omega-\omega_0)^2}{\Delta^2}\right]+\exp\!\left[-\frac{(\omega+\omega_0)^2}{\Delta^2}\right]\right\},
\label{e:G} \raisetag{-3pt}
\end{align}
and $E_0$, $\Delta$, and $\omega_0$ are the amplitude, width, and center of the electric field, respectively.
The phase mask produced using a pulse shaping technique\cite{Weiner1317,Weiner20113669} for 1D Raman spectroscopy is defined as
\begin{align}
\phi^{(3)}[\omega; \tau] = \alpha\cos[\tau(\omega-\omega_0)] ,
\label{e:phase mask1}
\end{align}
where $\tau$ is the time difference between the pulses and $\alpha=1.2$ is a mask constant chosen to adjust the pulse profile to the measurement. A vibrational excitation is then created by a sequence of pulses whose time period corresponds to a vibrational mode.\cite{Silberberg1} Thus, by sweeping $\tau$, we can tune the excitation frequency of the vibrational modes. 

The signal intensity is then evaluated from the electric field generated from the third-order polarization on the basis of a phase shift between the input and output electric fields arising from the coherent Stokes and anti-Stokes processes (see Appendix~\ref{s:Ramanprocess2})

\begin{figure*}[t!]
\includegraphics[width=0.8\textwidth]{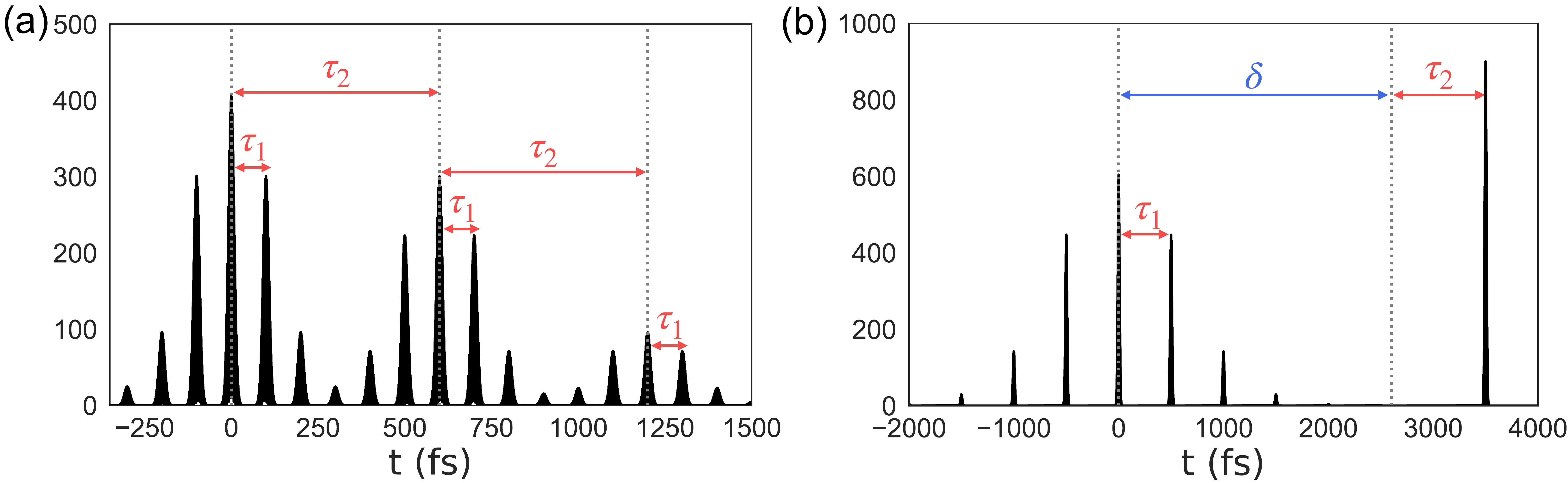}
\caption{\label{f:new_Et}Time-domain representation of the electric field used for SBSC 2D Raman spectroscopy for  liquid carbon tetrachloride: (a) original design using the phase mask given by Eq.~\eqref{e:phase mask}; (b)  new design given by Eq.~\eqref{e:5thnew_tau2}.} 
\end{figure*}

As demonstrated by Frostig \emph{et al.},\cite{Frostig} this SBSC stimulated Raman spectroscopy (SRS) can be extended to 2D Raman measurements. The fifth-order polarization is then expressed as (see Appendix~\ref{s:Ramanprocess3})
\begin{align}
P^{(5)}[\omega] 
={}& \left(\frac{1}{2\pi}\right)^2\int_{-\infty}^{\infty}{d\omega_1}\int_{-\infty}^{\infty}{d\omega_2}\,{R}^{(5)}[\omega_1,\omega_1+\omega_2]\nonumber\\
&\times{E'}[\omega-(\omega_1+\omega_2)]  
 \int_{-\infty}^{\infty}{d\Omega\,{E'}[\Omega]{E'}^*[\Omega-\omega_1]}\nonumber\\
 &\times\int_{-\infty}^{\infty}{d\Omega'\,{E'}[\Omega']{E'}^*(\Omega'-\omega_2]} , \label{e:5th_omega}
\end{align}
where $\Omega$ and $\Omega-\omega_j$ ($j=1, 2$) correspond to the pump and Stokes frequencies, respectively.  
The electric field ${E'}[\omega]$ that is originally employed in the fifth-order measurements is the same as in Eq.~\eqref{e:EwO},  apart from the phase mask, which is now defined as 
\begin{align}
E'[\omega] = G(\omega)  \exp(i\phi^{(5)}[\omega; \tau_1, \tau_2] ), 
\label{e:E1org} 
\end{align}
where 
\begin{align}
\phi^{(5)}[\omega; \tau_1, \tau_2] = \alpha\{\cos[\tau_{1}(\omega-\omega_0)]+\cos[\tau_{2}(\omega-\omega_0)]\},
\label{e:phase mask}\raisetag{-3pt}
\end{align}
and $\tau_{1}$ and $\tau_{2}$ correspond to the time delays between the pulse trains [see Fig.~\ref{f:new_Et}(a)]. These time delays determine the excitation frequencies  $\omega_1=1/\tau_{1}$ and $\omega_2=1/\tau_{2}$, respectively. Here, we chose ${E}[\omega] = {E}^*[-\omega]$ to set the time-domain electric field as a real function.  By exploiting  heterodyne detections, it is possible to separate out the undesired third-order cascading effects. \cite{cascade01,cascade02,cascade03}  Although the 2D Raman response function is asymmetric along the $\omega_1=\omega_2$ line,\cite{Inhomo01} the  2D signal obtained with this original mask design is symmetric, 
as can be seen from the definition in Eq.~\eqref{e:5th_omega}.\cite{couple04}  This symmetric feature of the signal makes  separation of the third- and fifth-order contributions to the 2D spectrum difficult.

Therefore, a new SBSC setup that consists of a spectrally shaped pump and a transform-limited probe has been developed.\cite{Hurwitz}  The electric field is now defined as
\begin{align}
E'[\omega] ={}& G(\omega)  \exp(i\phi^{(3)}[\omega;\tau_1]) \nonumber\\
&+ \mathcal{F}\!\!\left[E_0\cos[\omega'_0(t-\delta-\tau_2)]\exp\!\left(-\frac{(t-\delta-\tau_2)^2}{t_{\mathrm{FWHM}}^2}\right)\right], 
\label{e:5thnew_tau2}\raisetag{-6pt}
\end{align}
where $\omega'_0$, $\delta$, and $t_{\mathrm{FWHM}}$ are the central frequency,  delay time, and  width of the Gaussian envelope of the second pulse.  Because we have introduced the time delay $\tau_2$ in the frequency-domain measurement, the obtained 2D signal is not symmetric along $\omega_1=\omega_2$, which simplifies the signal analysis.  Moreover, because the excitation as a function of $\tau_2$ is not periodic [see Fig.~\ref{f:new_Et}(b)], we can suppress the artifact of overtone contributions, as is demonstrated below.

In both cases, the signal intensity is evaluated from the phase shift in a same manner as in the 1D Raman case,  illustrated in Appendix~\ref{s:Ramanprocess2}.

\section{Calculation methods}\label{secIII}

\subsection{MD simulations} \label{details:MD}

A detailed explanation of the MD simulation methods for 2D Raman spectroscopy is given in Ref.~\onlinecite{Hybrid02}. Here, we calculate the fifth-order response functions using an equilibrium--nonequilibrium hybrid MD simulation algorithm.\cite{Hybrid01} 

\subsubsection{Liquid carbon tetrachloride} 

We carried out MD simulations of liquid carbon tetrachloride using the optimized potentials for liquid simulations (OPLS) force field. Note that because the OPLS intramolecular interactions are harmonic in nature, any anharmonicity in the intramolecular modes is due solely to the intermolecular interactions. The simulations were performed under periodic boundary conditions, using 32 carbon tetrachloride molecules in a cubic simulation box of length 8.647~\AA, which corresponds to the experimental density of 1.58~g/cm$^3$.~\cite{Chang:1995} The equations of motion were integrated using the velocity-Verlet algorithm with a time steps of 2~fs. For the initial equilibration run, we performed isothermal $NVT$ simulations with a Nos\'e--Hoover thermostat at 298~K. We then carried out production runs using a hybrid MD approach. First, we  conducted equilibrium $NVE$ MD (EMD) simulations and then performed nonequilibrium MD (NEMD) simulations with a Raman laser field of 0.93~V/\AA. To account for  electrostatic interactions, we employed the direct reaction field (DRF) method,~\cite{Saito:2003bp} using the Ewald summation technique (with tinfoil boundary conditions) to compute the electronic polarizability of the system. The atomic polarizabilities of the carbon and chlorine atoms were taken to be 1.288599  and 2.40028~\AA$^3$, respectively, and were chosen such that the molecular polarizability matched the experimental value, 10.002~\AA$^3$.~\cite{Olney199759} The damping parameter in the DRF method for liquid  carbon tetrachloride was chosen to be 1.95163. 

\subsubsection{Liquid water} 
For the MD simulations of liquid water, we used the POLI2VS polarizable water model, designed for intermolecular and intramolecular vibrational spectroscopy.\cite{H-Water} We used a cubic simulation box, under periodic boundary conditions, containing 64 water molecules. The electrostatic interactions,  including the long-range charge--charge, charge--dipole, and dipole--dipole interactions, were again evaluated using a Ewald summation.\cite{2DIR} The interaction potentials containing the quadrupole moments were cut off at a distance equal to half the box size through the introduction of a smooth switching function. The equations of motion were integrated using the velocity-Verlet algorithm with ${\Delta}t=0.25$~fs. After the initial isothermal equilibration, the volume and energy of the system were fixed for both the EMD and NEMD simulations. The simulation parameters were chosen such that the average density and temperature were 0.997~g/cm$^3$ and 300~K, respectively. 

In both cases, the third-order response functions were calculated from the equilibrium trajectory data, while the fifth-order response functions were calculated using an equilibrium--nonequilibrium hybrid MD simulation algorithm.\cite{Hybrid01, Hybrid02}

\subsection{SBSC 1D and 2D Raman signals}
Using the response functions obtained from the MD simulations, we compute the third- and fifth-order polarizations from Eqs.~\eqref{e:3rd_omega} and~\eqref{e:5th_omega}, respectively. To evaluate the convolution integrals appearing in these expressions, we use the fast Fourier transform (FFT). The electric field generated from these polarizations is calculated from the Maxwell equation
\begin{align}
-\frac{\partial^2E(z,t)}{{\partial}z^2} + \frac{\epsilon^{(1)}}{c^2}\frac{\partial^2E(z,t)}{{\partial}t^2} = -\frac{1}{\epsilon_0c^2}\frac{\partial^2P^{(n)}(z,t)}{{\partial}t^2}, \label{e:waveeq}
\end{align}
where $E(z,t)$, $P^{(n)}(z,t)$, and $\epsilon^{(1)}$ are the generated electric field, the $n$th-order nonlinear polarization, and the relative permittivity (whose value depends on the material), respectively. The constants $\epsilon_0$ and $c$ are the vacuum permittivity and the speed of light. Under the tight focusing condition~\cite{tightfo} and the slowly varying amplitude approximation, the generated $n$th-order electric field in the frequency domain is expressed as~\cite{Hadasmaster}
\begin{align}
{E}_\mathrm{gen}^{(n)}[\omega] =
-\frac{i\omega}{2\epsilon_0n[\omega]c}{P}^{(n)}[\omega]\,dz, \label{e:Egenterated}
\end{align}
where $dz$ is the length of the volume element over which the electric field interacts with the sample molecules, and \(n[\omega]\) is the refractive index. We adjust $dz$ to fit the range of experimentally observed spectral shifts. The third- and fifth-order Raman spectra are calculated from  (see Fig.~\ref{f:Ewshift} in   Appendix~\ref{s:Ramanprocess2})
\begin{align}
I^{(3)}[\omega] &= \big|{E}_\mathrm{in}[\omega]+{E}^{(3)}_\mathrm{gen}[\omega]\big|^2, \label{e:I3} \\[6pt]
I^{(3+5)}[\omega] &= \big|{E}_\mathrm{in}[\omega]+{E}^{(3)}_\mathrm{gen}[\omega]+{E}^{(5)}_\mathrm{gen}[\omega]\big|^2, \label{e:I5}
\end{align}
respectively, where ${E}_\mathrm{in}[\omega]$ is the input electric field generated from the pulse shaper such that it is described by Eq.~\eqref{e:EwO}. It should be noted that the fifth-order signal defined by Eq.~\eqref{e:I5} includes the third-order contribution, which has to be eliminated to obtain the pure 2D Raman spectrum, using, for example,  sparse signal recovery techniques.\cite{Frostig_signalrecovery}  Here, we consider the case in which we can obtain $E_\mathrm{in}[\omega]+{E}^{(5)}_\mathrm{gen}[\omega]$ using the 1D Raman results for a given ${E}_\mathrm{in}[\omega]$. We then evaluate the pure 2D Raman signal as
\begin{align}
I^{(5)}[\omega] = \big|{E}_\mathrm{in}[\omega]+{E}^{(5)}_\mathrm{gen}[\omega]\big|^2. \label{e:I52}
\end{align}
The 1D and 2D Raman spectra are evaluated from the peak shift of the SRS signal that arises from the Raman gain of the Stokes peak and the Raman loss of the anti-Stokes peak (see Appendix~\ref{s:Ramanprocess2}).

Here, the phase shifts between the input and the generated electric field are evaluated from the spectral centroid functions, defined as
\begin{align}
\nu_\mathrm{cent}^{(3)}(\tau) &= \frac{\displaystyle\int_{a}^{b}d\omega\,{I^{(3)}[\omega;\tau]\omega}}{\displaystyle\int_{a}^{b}d\omega\,{I^{(3)}[\omega;\tau]}}, \label{e:3centroid} \\[6pt]
\nu_\mathrm{cent}^{(5)}(\tau_1,\tau_2) &= \frac{\displaystyle\int_{a}^{b}d\omega\,{I^{(5)}[\omega;\tau_1,\tau_2]\omega}}{\displaystyle\int_{a}^{b}d\omega\,{I^{(5)}[\omega;\tau_1,\tau_2]}}, \label{e:5centroid} 
\end{align}
where $a$ and $b$ are the lower and upper limits of the spectral peak. The SBSC 1D and 2D Raman spectra are obtained 
by Fourier transforming the above functions as $\nu_\mathrm{cent}^{(3)}[\omega]$ and $\nu_\mathrm{cent}^{(5)}[\omega_1,\omega_2]$, respectively.

The procedure for calculating SBSC 1D and 2D Raman spectra can be outlined as follows. 
\begin{enumerate}
\item Compute the third- and fifth-order response functions $R^{(3)}(t_1)$ and $R^{(5)}(t_1,t_2)$ from the MD simulations. 
\item For 1D spectra, calculate the third-order polarization $P^{(3)}[\omega]$ from Eq.~\eqref{e:3rd_omega}, using the electric field $E[\omega]$  defined by Eqs.~\eqref{e:EwO}--\eqref{e:phase mask1}. For the 2D spectra, calculate the fifth-order polarization $P^{(5)}[\omega]$ using Eq.~\eqref{e:5th_omega} with    Eqs.~\eqref{e:E1org}  and~\eqref{e:phase mask} for the original design, and Eq.~\eqref{e:5th_omega} with Eq.~\eqref{e:5thnew_tau2} for the improved design. 
\item Calculate  from Eq.~\eqref{e:Egenterated} the electric fields generated from the third- and fifth-order polarizations $E^{(3)}[\omega]$ and $E^{(5)}[\omega]$.
\item Evaluate the intensities of the third- and fifth-order Raman signals $I^{(3)}[\omega]$ and $I^{(5)}[\omega]$ from Eqs.~\eqref{e:I3} and~\eqref{e:I52}. 
\item Calculate the centroid functions $\nu_\mathrm{cent}^{(3)}(\tau)$ and $\nu_\mathrm{cent}^{(5)}(\tau_1,\tau_2)$ from  Eqs.~\eqref{e:3centroid} and~\eqref{e:5centroid}. Repeat steps~2--4 for different $\tau$  for 1D spectra, and $\tau_1$ and $\tau_2$  for 2D spectra.  The SBSC 1D and 2D Raman spectra are obtained by Fourier transforming the centroid functions as $\nu_\mathrm{cent}^{(3)}[\omega]$ and $\nu_\mathrm{cent}^{(5)}[\omega_1,\omega_2]$, respectively.
\end{enumerate}

\begin{figure*}[t!]
\includegraphics[width=0.8\textwidth]{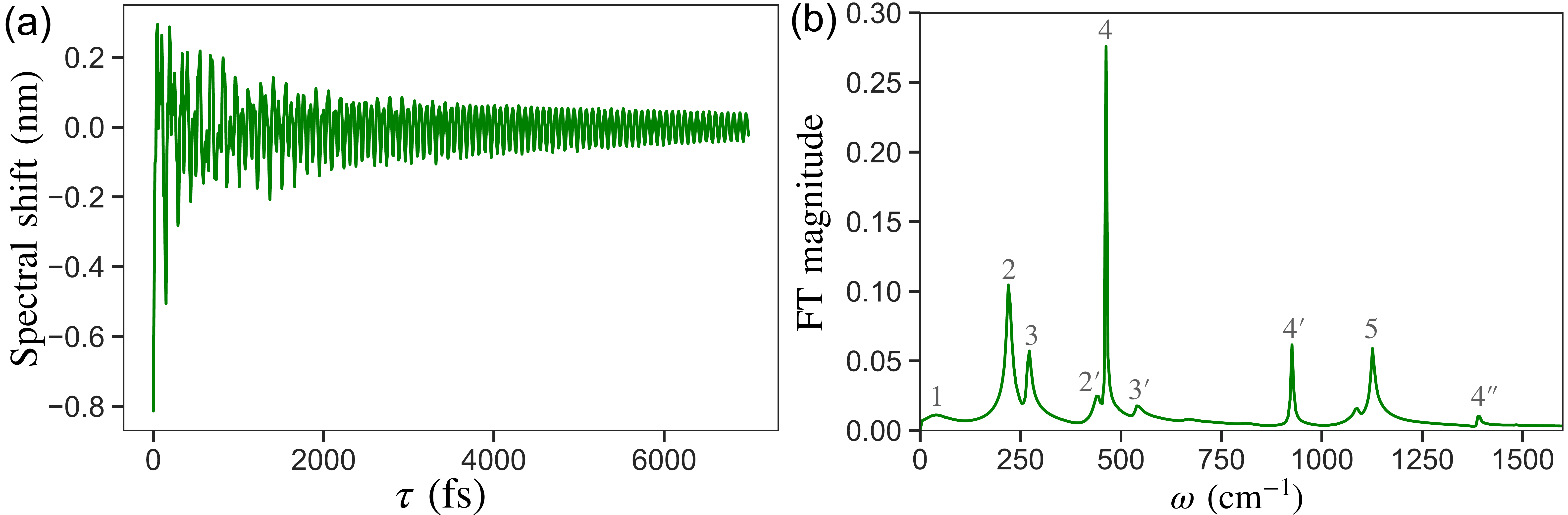}
\caption{\label{f:CCl4_1d}(a) Spectral shifts of liquid carbon tetrachloride calculated from the MD simulations for the third-order response function. We computed the spectral shifts in the range  ($-1$~nm,~1~nm), adjusting the parameter values in Eq.~\eqref{e:Egenterated}  as reported experimentally. (b)  SBSC 1D Raman spectrum $\nu_\mathrm{cent}^{(3)}[\omega]$} of liquid carbon tetrachloride calculated from the spectral shifts. The five peaks, labeled $n=1$ to $5$, correspond to the inter- and intramolecular vibrational modes of liquid carbon tetrachloride, while the $n'=2'$ to $4'$ and  $n''=4''$ peaks are  artifacts of the SBSC measurement.
\end{figure*}

\section{Results and discussion}\label{secIV}

\subsection{SBSC 1D Raman spectra}

\subsubsection{Liquid carbon tetrachloride}

In Fig.~\ref{f:CCl4_1d}, we display the SBSC 1D Raman signal of liquid carbon tetrachloride obtained from the MD simulation using the procedure described in Sec.~\ref{secIII}. Here, we set the width and the central frequency of the Gaussian electric field as $\Delta=3335$~cm$^{-1}$ and $\omega_0= 12\,578$~cm$^{-1}$, and we choose $E_0$   to be the same as in the MD simulation. The conventional 1D Raman spectrum calculated using the same force field under the same conditions was presented in Ref.~\onlinecite{JoCP2016}.
The first peak, labeled  $n=1$, arises from the intermolecular vibrational modes. The remaining peaks, labeled  $n=2$ to $5$, arise from the intramolecular vibrational modes of liquid carbon tetrachloride: they correspond to the $\nu_2$ bending, $\nu_4$ bending, $\nu_1$ symmetric stretching, and $\nu_3$ stretching motions of the molecule, respectively. The resonant frequencies of peaks 1, 2, and 4 are in good agreement with the experimental result. Peaks $3$ and $5$ are shifted toward the red and blue from the experimental results by 40 and 345~cm$^{-1}$, respectively, because the OPLS force field does not properly account for the resonant frequencies of the intramolecular modes.

As explained in Appendix~\ref{s:Ramanprocess1}, the Raman transitions created by the SBSC measurement are not only $\omega=1/\tau$, but also $\omega=1/2\tau$ and $\omega=1/3\tau$, arising from every one, two, and three pulse excitations in the pulse train, where $\tau$ is the time duration between  pulses. Thus, when we measure the system with the resonant peak at $\omega=\nu_n$ using this spectroscopic method, we also observe the spurious peaks at $\omega' =2\nu_n$ and $\omega''=3\nu_n$. In Fig.~\ref{f:CCl4_1d}, these peaks are labeled  $n'=2'$ to $4'$ and  $n''=4''$.

\subsubsection{Liquid water}

\begin{figure*}[t!]
\includegraphics[width=0.8\textwidth]{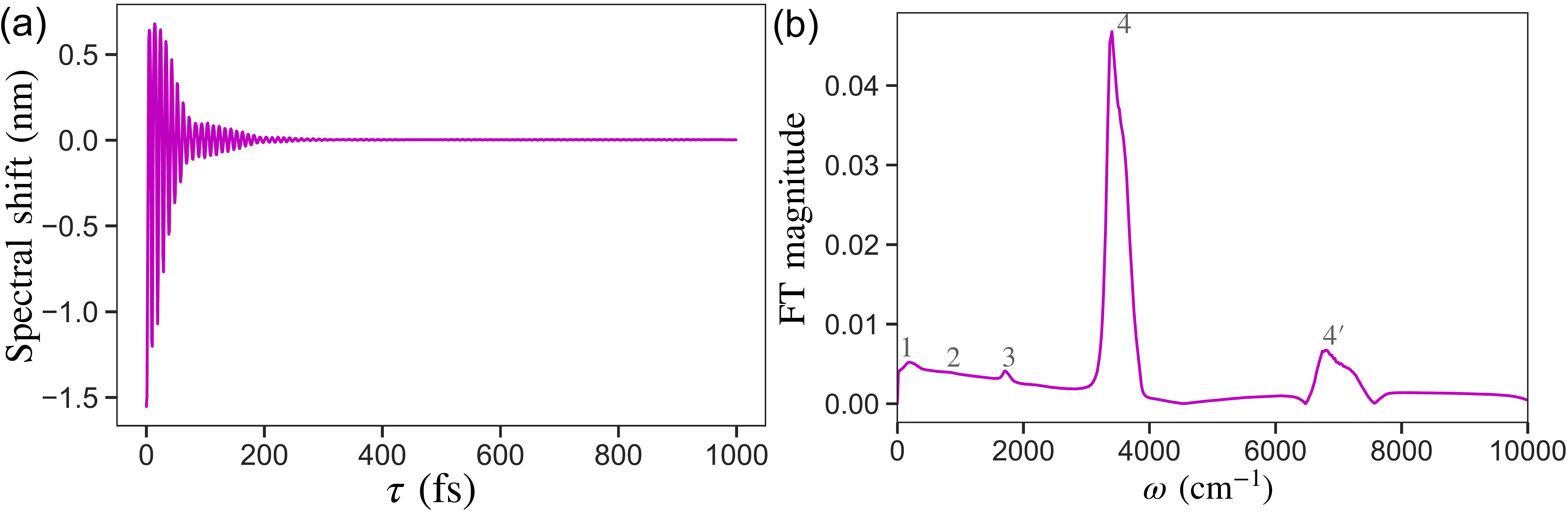}
\caption{\label{f:H2O_1d}(a) 
Spectral shifts of liquid water calculated from the MD simulations for the third-order response function.  We computed the spectral shifts in the range  ($-1$~nm,~$1$~nm), adjusting the parameter values in Eq.~\eqref{e:Egenterated} as reported experimentally. (b)  SBSC 1D Raman spectrum $\nu_\mathrm{cent}^{(3)}[\omega]$ for liquid water obtained from the spectral shift.  The four peaks labeled  $n=1$ to $4$  correspond to the hydrogen bond (HB) intermolecular vibrational,  HB intermolecular librational, intramolecular H--O--H bending, and intramolecular O--H stretching modes, respectively,\cite{2DIR} whereas the broadened background signal from the 0--3000~cm$^{-1}$ region and the  peak labeled $n'=4'$ are  artifacts of the SBSC measurement.}
\end{figure*}

In Fig.~\ref{f:H2O_1d}, we display the SBSC 1D Raman spectrum of liquid water calculated from the MD simulations.   Here, we set the width and the central frequency of the Gaussian electric field as $\Delta= 11\,118~$cm$^{-1}$ and $\omega_0= 23\,148$cm$^{-1}$, and we choose $E_0$   to be the same as in the MD simulation. We choose a much larger $\Delta$ here than in the case of carbon tetrachloride in order to suppress the range of the pulse train, because the vibrational motion of liquid water decays very rapidly [see Fig.~\ref{f:H2O_1d}(a)]. 
We have reported the conventional 1D Raman spectrum, using the same force field under the same conditions, in Refs.~\onlinecite{2DIR,H-Water, JoCP2016}. These results indicated that peaks 1, 3, and 4 correspond to the hydrogen bond (HB) intermolecular,  H--O--H bending, and  O--H stretching modes, respectively. The tiny peak labeled  2 corresponds to the HB intermolecular librational mode.  Because the intermolecular charge transfer effect is not properly included in the POLI2VS potential, the peak intensity associated with this mode is underestimated.\cite{IHTpol} The H--O--H bending and O--H stretching peaks appear around  1700 and 3350~cm$^{-1}$, which are higher than the experimentally obtained peak positions. This is because we have calculated the signals in the classical limit, whereas   POLI2VS was developed for quantum mechanical calculations.\cite{JJLiu18,Sakurai-JPCA-2011-115}

As in the case of carbon tetrachloride, we observe  spurious peaks as  artifacts of the SBSC measurements. Thus, the peak labeled $4^\prime$, located around 7000~cm$^{-1}$, appears as the spurious peak of the O--H stretching peak at 3500~cm$^{-1}$. Moreover, we observe a broad background spectral peak from 0 to 3000~cm$^{-1}$, which is due to the contributions of the spurious peaks $n'=1'$ and $2'$, and  $n''=1''$ and $2''$, which merge with the original peaks $n=1$ and $2$. 

 As well as these peaks, we find that the peak positions and the ratio of peak intensity among $n=1$ to $4$ agree with the conventional 1D Raman results calculated under the same conditions:\cite{H-Water, JoCP2016} the SBSC 1D Raman measurement captures the essential features of the intermolecular and intramolecular motion of water, as in the conventional 1D Raman spectrum.

\subsection{SBSC 2D Raman spectra}
\subsubsection{Liquid carbon tetrachloride}

\begin{figure*}[t!]
\includegraphics[width=0.8\textwidth]{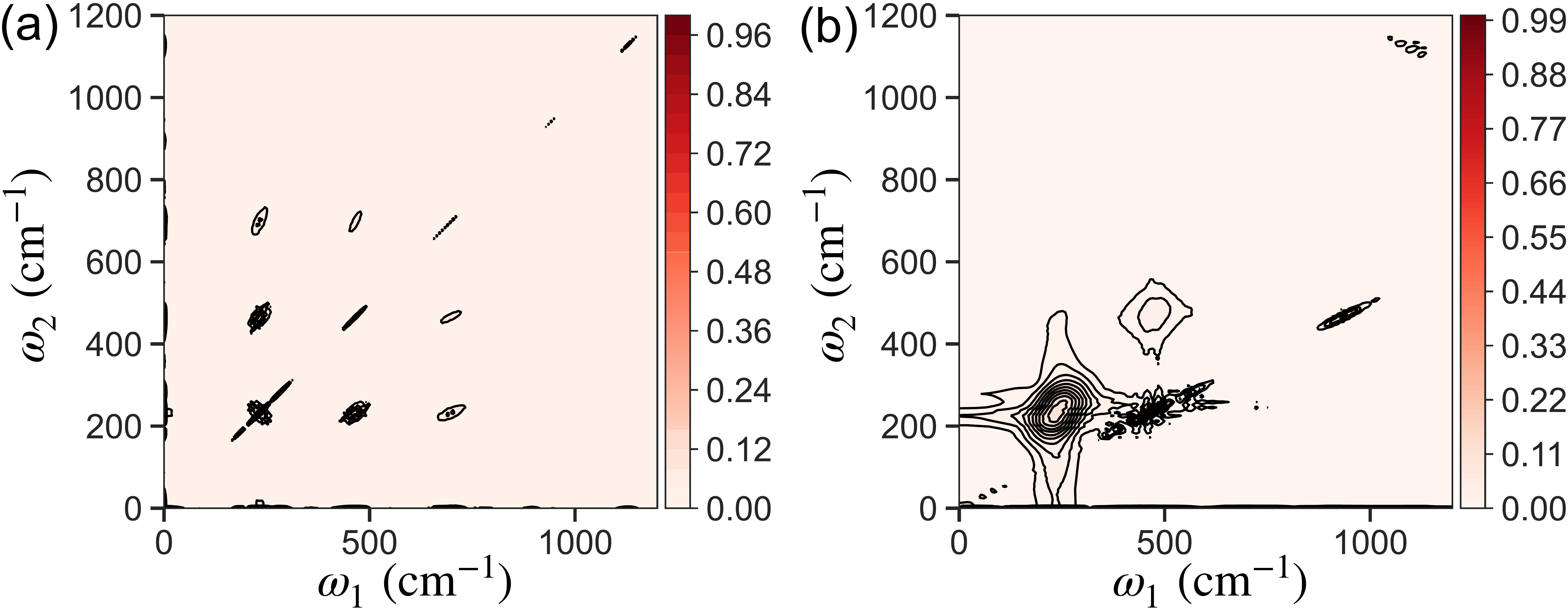}
\caption{\label{f:2dresponse} SBSC 2D Raman spectra $\nu_\mathrm{cent}^{(5)}[\omega_1,\omega_2]$} for liquid carbon tetrachloride obtained from the MD simulations for (a) the original  and (b) the improved electric fields.
\end{figure*}

In Fig~\ref{f:2dresponse},  we present the simulation results of SBSC 2D Raman spectra for liquid carbon tetrachloride calculated using (a) the original electric field defined in Eq.~\eqref{e:E1org}  and (b) the improved electric field defined in Eq.~\eqref{e:5thnew_tau2}.  The parameter values of the Gaussian envelope were chosen to be the same as in  Fig.~\ref{f:CCl4_1d}.  For the improved pulse design, we further set $\delta=2600$~fs and $\omega'_0= 18\,796$~cm$^{-1}$.  The conventional 2D Raman signal calculated using the same force field has been presented in Ref.~\onlinecite{JoCP2016}.

In Fig.~\ref{f:2dresponse}(a), because the phase mask given by Eq.~\eqref{e:phase mask} is symmetric in $\tau_1$ and $\tau_2$, the 2D spectrum obtained from the original electric field is symmetric along the $\omega_1=\omega_2$ line. In this case, the spurious peaks for the double or triple resonant frequencies appear not only at the off-diagonal positions  $(\omega_1,\omega_2) = (\omega_2,\omega_1) = (446,223)$, $(544, 272)$, $(669,223)$, and $(816,272)$, but also along the diagonal  at $(926, 926)$ (all in units of cm$^{-1}$). Although the conventional 2D Raman spectrum does not usually exhibit  peaks along the $\omega_1=\omega_2$ line,\cite{JoCP2016}  here we observe  diagonal peaks at 223, 272, 463, and 1125~cm$^{-1}$. This is because the phase mask given by Eq.~\eqref{e:phase mask}  becomes identical to that of the 1D case in Eq.~\eqref{e:phase mask1}  for $\tau_1=\tau_2$,  except the prefactor  $2\alpha$, and thus the spectrum along the $\omega_1=\omega_2$ line becomes similar to the 1D Raman spectrum presented in Fig.~\ref{f:CCl4_1d}(b).  Moreover, along the $\omega_1$ axis (or the $\omega_2$ axis) in Fig.~\ref{f:2dresponse}(a), we observe the peaks at similar positions to those in the 1D case, because the spectrum along this axis is measured after the effect of the first excitation has vanished for $\tau_2\approx 2\pi/\omega_2 \gg 2\pi/\nu_1$, where $\nu_1$ is the vibrational frequency of the slowest mode. 

In Fig~\ref{f:2dresponse}(b), with the improved phase mask, the number of spurious peaks appearing in the spectrum is dramatically reduced, because the excitation in the $\tau_2$ direction is created only by a single pulse [see Fig.~\ref{f:new_Et}(b)],  and thus the double and triple quasi-excitations in the $\omega_2$ direction are suppressed: these peaks appear only below the diagonal direction.  Thus, using the improved electric field,  we can clearly observe the mode--mode coupling peaks at $(\omega_1, \omega_2)=(463, 223)$ and $(\omega_1, \omega_2)=(463, 272)$, while the peak at $(\omega_1, \omega_2)=(926,463)$ is the spurious peak of the $(463,463)$ peak in the $\omega_1$ direction.

\subsubsection{Liquid water}

\begin{figure*}[t!]
\includegraphics[width=0.8\textwidth]{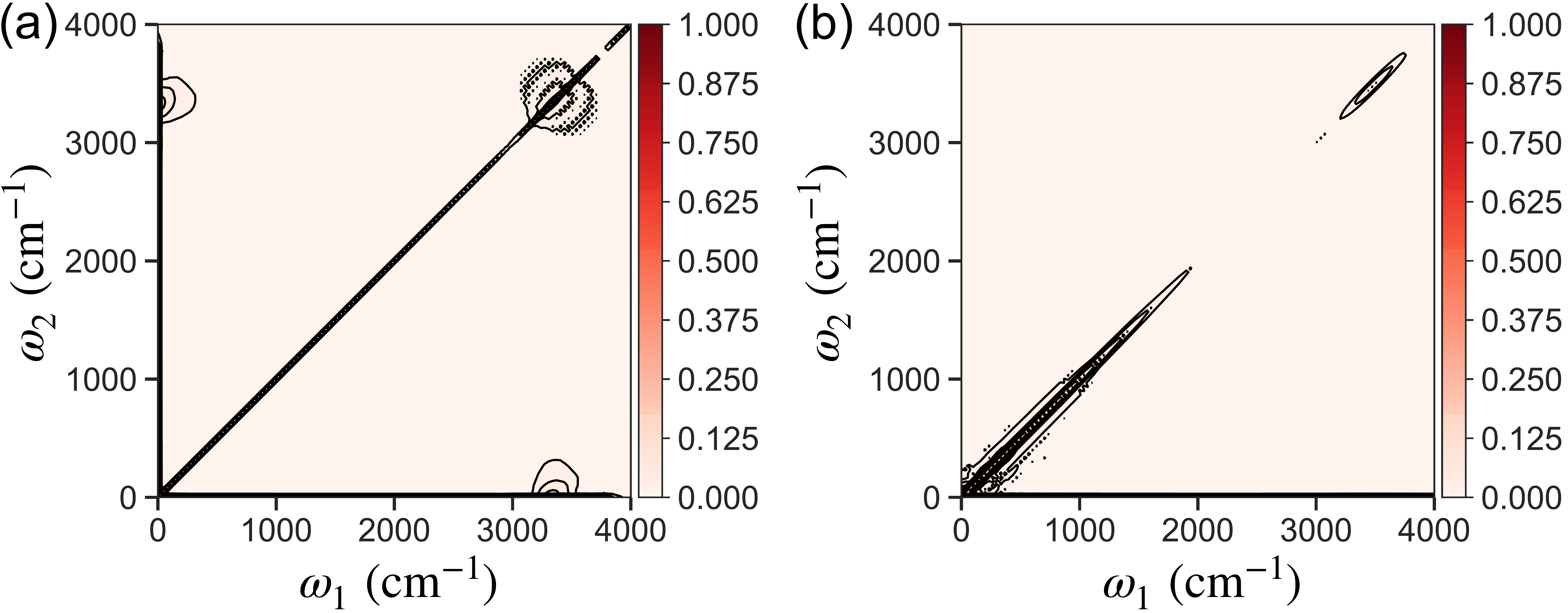}
\caption{\label{f:water2dresponse} SBSC 2D Raman spectra $\nu_\mathrm{cent}^{(5)}[\omega_1,\omega_2]$ for water obtained from the MD simulations for (a) the original and (b) the improved electric fields.}
\end{figure*}

In Fig.~\ref{f:water2dresponse}, we present the simulation results of the SBSC 2D Raman spectrum for liquid water calculated using (a) the original electric field  and (b) the improved electric field.  The parameter values of the Gaussian field were chosen to be the same as in the 1D case for water.  For the improved pulse design, we set $\delta=100$~fs, $\Delta = 11\,118$~cm$^{-1}$, $t_{\mathrm{FWHM}} = 10$~fs ($=3335$~cm$^{-1}$), $\omega_0 = 25\,000$~cm$^{-1}$, and $\omega'_0= 9090$~cm$^{-1}$.  We choose  small $\delta$ and $t_{\mathrm{FWHM}}$ here because the vibrational motion of liquid water decays very rapidly [see Fig.~\ref{f:H2O_1d}(a)], while we wish to maintain the spectral resolution for the peaks near $3500$~cm$^{-1}$. 
A detailed analysis of the conventional 2D Raman signal of water, on the basis of MD simulations and a Brownian model, has been presented in the context of 2D IR-Raman spectroscopy in Ref.~\onlinecite{2DIR}. It should be noted that because the intermolecular charge transfer effect is not properly included in the POLI2VS force field, the present descriptions of 2D Raman spectra in the low-frequency intermolecular-mode range $0 \le \omega_1 \le 1000$~cm$^{-1}$ and/or $0 \le \omega_2 \le1000$~cm$^{-1}$  may not be accurate.\cite{IHTpol}

In  Fig~\ref{f:water2dresponse}(a), we observe the signal along the $\omega_1$ and $\omega_2$ lines,
as in the case of carbon tetrachloride in Fig~\ref{f:2dresponse}(a). Although we observe a prominent peak at 3500~cm$^{-1}$ representing the O--H stretching modes, the signal below 3000~cm$^{-1}$ is featureless, because the contributions of the spurious peaks with double and triple frequencies, originating from the inhomogeneously broadened low-frequency peaks, are overlapped and  mixed. 
From the analysis of the 2DIR-Raman signal, the peak observed around $(\omega_1,\omega_2) = (50,3300)$ in Fig.~\ref{f:water2dresponse}(a)  can be regarded as a cross-peak arising from   intermolecular and intramolecular mode--mode coupling.  
Although the profile of this cross-peak is much sharper than that  obtained from the simulation of the conventional 2D Raman measurement,  this is because the SBSC measurement is conducted in a very different manner from  conventional spectroscopy:  the excitation is created by a train of pulses, and the signal intensity is evaluated indirectly on the basis of the spectral phase shift in the emission signal.

In the modified electric field case in Fig~\ref{f:water2dresponse}(b), the presence of the spurious peaks along the $\omega_1=\omega_2$ line is suppressed, 
while the appearance of the O--H stretching peak at $(\omega_1,\omega_2) = (3500,3500)$ becomes clear. A signal profile in the low-frequency region that seems to involve  contributions from  intermolecular mode--mode coupling is also revealed. 
 The signal along the axis appears only in the $\omega_1$ direction, because  creation of the spurious peaks is suppressed in the $\omega_2$  direction.  On the contrary, we cannot observe a mode--mode coupling peak at $(\omega_1,\omega_2) = (50,3300)$, because the excitation in the $\omega_2$ direction in this case is created by a single pulse [see Fig.~\ref{f:new_Et}(b)] and is not mode-selective, in contrast to the original electric field using a  pulse train [see Fig.~\ref{f:new_Et}(a)]. 
This indicates that the original electric field is advantageous for the detection of mode--mode coupling peaks with a large frequency difference, while the asymmetric 2D spectrum obtained using the modified electric field can provide  useful information to identify the separate contributions of the signals from the $\omega_1$ and $\omega_2$ directions. 

Here, we chose the  parameter values of the electric field to focus on the HB stretching peak at about  3500~cm$^{-1}$. If we wish to observe the intermolecular modes more clearly, however, we need to design a phase mask specifically for  low-frequency modes. To observe  mode--mode coupling peaks between  high-frequency intramolecular  and low-frequency intermolecular modes, we need to design the phase masks in the $\tau_1$ and $\tau_2$ directions in a different manner.

\section{Conclusion}\label{secV}
A recently developed SBSC 2D Raman spectroscopic technique has created new possibilities for measuring complex molecular interactions in condensed phases. In the present work, we have illustrated the key features of this technique and have described a method for computing SBSC Raman spectra on the basis of  MD simulations. 
Although analysis of these 2D measurements is not straightforward, because the 2D Raman signal is described by a second-order response function in which the both anharmonic mode--mode coupling and nonlinear polarizability play a significant role, we can obtain pure 2D Raman signals by eliminating the undesired cascading contributions that constitute the major difficulty in conventional 2D Raman spectroscopy. 

Using  simulation results of SBSC 2D Raman spectra for liquid carbon tetrachloride and liquid water, we have demonstrated the capabilities of the new technique for detecting not only low-frequency intermolecular modes but also  high-frequency intramolecular modes,  including their mode--mode coupling peaks. This single-shot  measurement technique has the capability of measuring the nonlinear Raman responses of  complex materials covering  wavelengths from the terahertz to the infrared region using a low-energy laser with a deep penetration depth in the sample.  Moreover, this SBSC 2D Raman technique can be implemented for imaging, and specifically in microscopy, making it possible to carry out spatially resolved measurements of molecular structure in biological or nanomaterial samples. Analysis of spectra obtained using this technique is not simple, however, because the excitation and detection of a system are performed by a laser in a very different way from that used in  conventional laser spectroscopy: the excitation of the molecular system is  by  a train of pulses, and signal detection   is  performed indirectly through the phase shift arising from Stokes and anti-Stokes Raman processes. Thus, the theoretical guidance provided by  MD simulations is important. 

To make a direct comparison between the results of our simulations  and those obtained experimentally, however, we must improve the force field and polarization function in accordance with those available in experimental systems.  Moreover, we need to carry out quantum dynamical simulations, in particular for intramolecular modes.\cite{JJLiu18,Sakurai-JPCA-2011-115}  
Nevertheless, we believe that the present results elucidate the key features of SBSC 2D Raman spectroscopic methods with regard to probing the fundamental nature of intermolecular and intramolecular interactions. Further investigations to find a systematic way to identify the peak profile are also necessary to foster the development of this spectroscopic method from a theoretical point of view. 
Using the simulation methods described in this paper, we can provide valuable information that can be applied to these 2D Raman experiments, allowing them to be carried out more efficiently.

\begin{acknowledgments}
J. Jo acknowledges  stimulating discussions with Hadas Frostig and Ilan Hurwitz. 
J. Jo was supported by  Japanese Government (MEXT)  Scholarships. The authors thank Hironobu Ito for providing simulation data on liquid water required for our calculations. 

\end{acknowledgments}

\section*{Data Availability}
The data that support the findings of this study are available from the corresponding authors upon reasonable request.

\begin{figure*}[t!]
\includegraphics[width=0.8\textwidth]{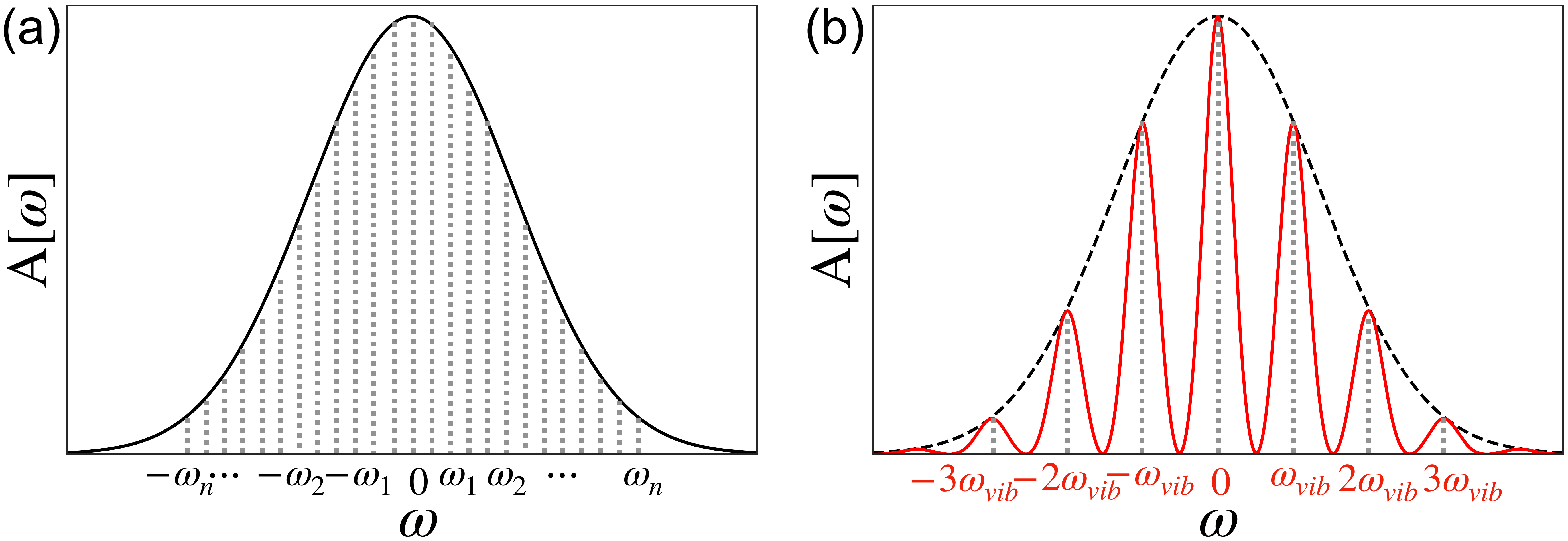}
\caption{\label{f:Aw} Transform-limited electric field (black) and  phase-modulated electric field (red). The phase modulation is produced with a cosine function with period $\omega_\mathrm{vib}$. 
For the transform-limited pulse (a), all vibrational modes within the range $- \omega_n\le \omega \le \omega_n $ are excited, whereas  for the phase-modulated pulse (b), vibrational modes with $\pm n \omega_\mathrm{vib}$ are selectively excited. }
\end{figure*}

\begin{figure}[t!]
\includegraphics[width=0.9\columnwidth]{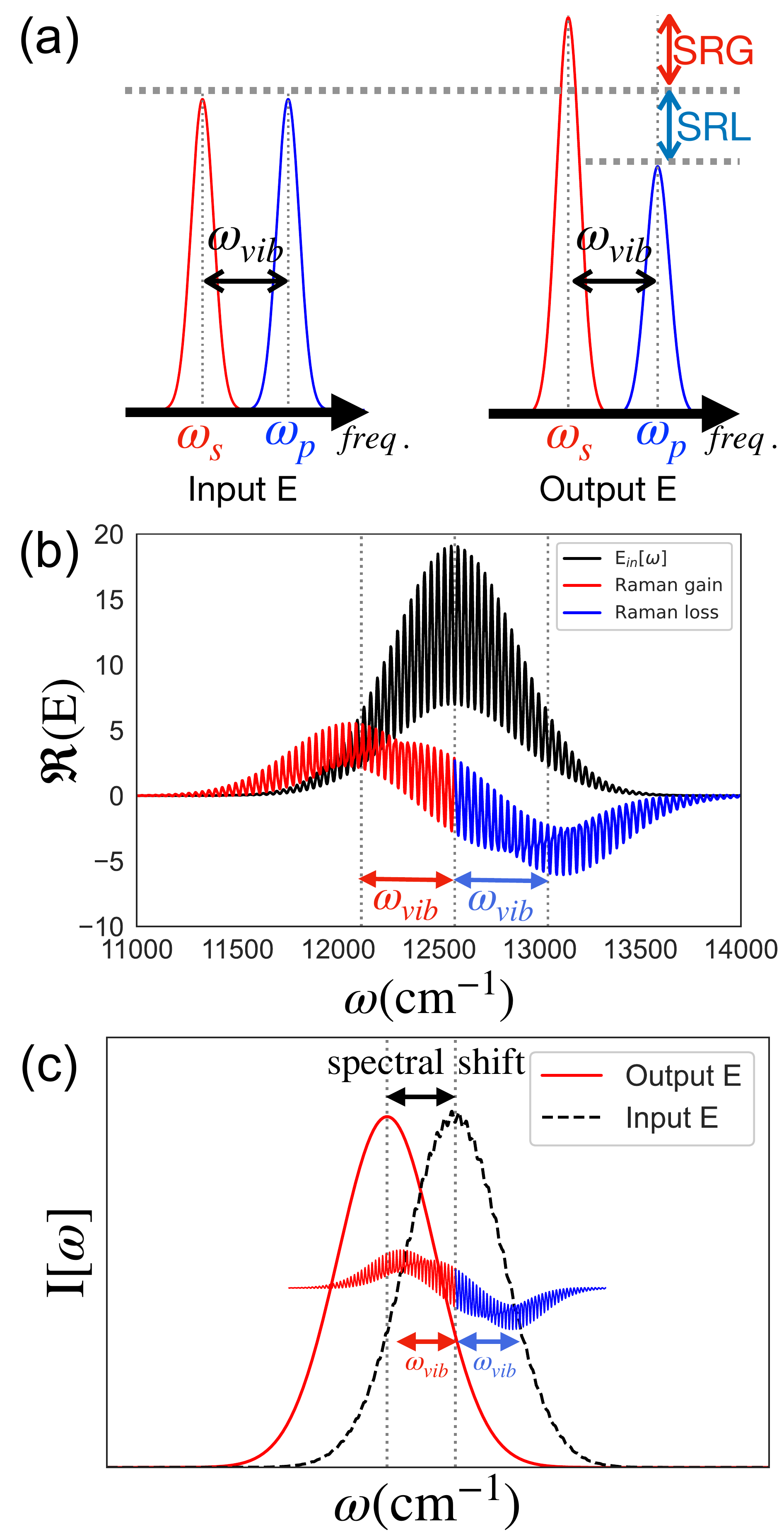}
\caption{\label{f:Ewshift} Schematic illustration of spectral phase shift.(a) In a multi-beam SRS case, when the difference between the pump pulse and the Stokes pulse matches a vibrational mode of the molecule $\omega_\mathrm{vib}$, stimulated excitation of Raman transitions occurs. The pump pulse exhibits stimulated Raman gain (SRG), depicted by the red curve, whereas the Stokes pulse exhibits stimulated Raman loss (SRL) depicted by the blue curve. (b) In a single-beam SRS case, the real part of the input electric field (black curve) and the generated electric field in the third-order polarization (red and blue curves). A Raman loss and a Raman gain occur in the signal at  frequencies $\omega_0+\omega_\mathrm{vib}$ and $\omega_0-\omega_\mathrm{vib}$, respectively, with $\omega_0$ being the central frequency of the pulse. Here, the generated electric field has been amplified by a factor of 20 to aid in visualization. (c) Because we use a single broadband pulse, the Raman loss and gain in the single-beam Raman spectroscopy result in a redshift.}
\end{figure}

\appendix
\section{1D Raman excitation}\label{s:Ramanprocess1}
The third-order polarization function in Eq.~\eqref{e:3rd_omega} is rewritten as
\begin{align}
P^{(3)}[\omega] 
=\frac{1}{2\pi}\int_{-\infty}^{\infty}{{d\omega_1}\,
{R}^{(3)}[\omega_1]{E}[\omega-\omega_1]} A[\omega_1],   
\label{e:3rd_appendix2}
\end{align}
where $E[\omega]$ is the Fourier representation of $E(t)$ and
\begin{align}
A[\omega] 
= \int_{-\infty}^{\infty} {{d\Omega}\,{E}[\Omega]{E}^*[\Omega-\omega]} \label{e:excitationAw}
\end{align} 
is the probability amplitude. The excitation of the vibrational modes in the molecular system is proportional to the probability amplitude. 

We now consider an electric field with phase $\phi[\omega]$  defined as (see Fig.~\ref{f:pshaper})
\begin{align}
E[\omega] = \big|E[\omega]\big|e^{i\phi[\omega]}. \label{e:excitationEw}
\end{align}
The probability amplitude is then expressed as (see Fig.~\ref{f:Aw})
\begin{align}
A[\omega] 
= \int_{-\infty}^{\infty}{{d\Omega}\,\big|{E}[\Omega]\big|\,\big|{E}[\Omega-\omega]\big|e^{i(\phi[\Omega]-\phi[\Omega-\omega])}}, \label{e:excitationAw2}
\end{align}
and  is therefore maximum for $\phi[\Omega]=\phi[\Omega-\omega]$. For a transform-limited pulse, such as the Gaussian pulse $G(\omega)$ in Eq.~\eqref{e:G} without a phase mask, the electric field excites all vibrational modes of a molecular system within the frequency range $- \Delta \le \omega \le \Delta $ [see Fig.~\ref{f:Aw}(a)].  In this case, all vibrational modes in this frequency range are excited simultaneously, and we cannot have a spectral resolution less than $2\Delta$.  On the contrary, if we introduce a phase mask and set $\phi[\Omega]$ to be a periodic function with a period $\omega_\mathrm{vib}= 2 \pi/\tau$, we can selectively excite the molecular system at $\omega=\pm \omega_\mathrm{vib}$, for which  $\phi[\Omega]=\phi[\Omega-\omega_\mathrm{vib}]$ is satisfied.  

We note, however, that because a pulse train with  time duration $\tau'=\tau/n$, where $n=2, 3, \dots$, can also excite the same mode $\omega=\omega_\mathrm{vib}$ with the participation of every $n$ pulses in the $\tau'$ pulse train, we also observe, as  artifacts of this measurement, spurious peaks of the mode $\omega_\mathrm{vib}$ at 
$\omega=n\omega_\mathrm{vib}= 2 \pi/\tau'$ that also satisfy the condition $\phi[\Omega]=\phi[\Omega-n\omega_\mathrm{vib}]$.

\section{Optical detection}\label{s:Ramanprocess2}
The polarization induced by this electric field causes a Raman loss  and a Raman gain of the signal at  frequencies $\omega_0+\omega_\mathrm{vib}$ and $\omega_0-\omega_\mathrm{vib}$, respectively, where $\omega_0$ is the central frequency of the pulse. When we use heterodyne detection with a single broadband pulse as a local oscillator, the Raman loss and gain result in a red frequency shift, as illustrated in Figs.~\ref{f:Ewshift}(b) and \ref{f:Ewshift}(c). The amplitude of this shift is proportional to the Raman intensity.\cite{hadasmaster2, CARS}  Thus, by measuring the spectral shift of the output field as a function of $\tau$ (the phase modulation period), we can evaluate the intensity of the SBSC spectrum.

\section{2D Raman excitation}\label{s:Ramanprocess3}
The extension to  2D Raman measurements is straightforward. The fifth-order polarization function in Eq.~\eqref {e:5th_omega} is now expressed as
\begin{align}
P^{(5)}[\omega] 
= {}&\left(\frac{1}{2\pi}\right)^2\int_{-\infty}^{\infty}{d\omega_1}\int_{-\infty}^{\infty}{d\omega_2}\,{R}^{(5)}[\omega_1,\omega_1+\omega_2]\nonumber\\
&\times{E'}[\omega-(\omega_1+\omega_2)] 
A[\omega_1] A[\omega_2]. 
\label{e:5th_appendix}
\end{align}
For this expression, the input electric fields in the time-domain representation for different pulse designs, Eqs. \eqref{e:E1org} and~\eqref{e:5thnew_tau2}, are depicted in Figs.~\ref{f:new_Et}(a) and~\ref{f:new_Et}(b), respectively. The detection of the signal can be conducted in the same way as explained in Appendix~\ref{s:Ramanprocess2}.
\clearpage

\section*{REFERENCES}

\bibliography{library}

\begin{thebibliography}{67}%
\makeatletter
\providecommand \@ifxundefined [1]{%
 \@ifx{#1\undefined}
}%
\providecommand \@ifnum [1]{%
 \ifnum #1\expandafter \@firstoftwo
 \else \expandafter \@secondoftwo
 \fi
}%
\providecommand \@ifx [1]{%
 \ifx #1\expandafter \@firstoftwo
 \else \expandafter \@secondoftwo
 \fi
}%
\providecommand \natexlab [1]{#1}%
\providecommand \enquote  [1]{``#1''}%
\providecommand \bibnamefont  [1]{#1}%
\providecommand \bibfnamefont [1]{#1}%
\providecommand \citenamefont [1]{#1}%
\providecommand \href@noop [0]{\@secondoftwo}%
\providecommand \href [0]{\begingroup \@sanitize@url \@href}%
\providecommand \@href[1]{\@@startlink{#1}\@@href}%
\providecommand \@@href[1]{\endgroup#1\@@endlink}%
\providecommand \@sanitize@url [0]{\catcode `\\12\catcode `\$12\catcode
  `\&12\catcode `\#12\catcode `\^12\catcode `\_12\catcode `\%12\relax}%
\providecommand \@@startlink[1]{}%
\providecommand \@@endlink[0]{}%
\providecommand \url  [0]{\begingroup\@sanitize@url \@url }%
\providecommand \@url [1]{\endgroup\@href {#1}{\urlprefix }}%
\providecommand \urlprefix  [0]{URL }%
\providecommand \Eprint [0]{\href }%
\providecommand \doibase [0]{http://dx.doi.org/}%
\providecommand \selectlanguage [0]{\@gobble}%
\providecommand \bibinfo  [0]{\@secondoftwo}%
\providecommand \bibfield  [0]{\@secondoftwo}%
\providecommand \translation [1]{[#1]}%
\providecommand \BibitemOpen [0]{}%
\providecommand \bibitemStop [0]{}%
\providecommand \bibitemNoStop [0]{.\EOS\space}%
\providecommand \EOS [0]{\spacefactor3000\relax}%
\providecommand \BibitemShut  [1]{\csname bibitem#1\endcsname}%
\let\auto@bib@innerbib\@empty
\bibitem [{\citenamefont {Mukamel}(1999)}]{Mukamel}%
  \BibitemOpen
  \bibfield  {author} {\bibinfo {author} {\bibfnamefont {S.}~\bibnamefont
  {Mukamel}},\ }\href@noop {} {\emph {\bibinfo {title} {{Principles of
  Nonlinear Optical Spectroscopy}}}}\ (\bibinfo  {publisher} {Oxford University
  Press},\ \bibinfo {year} {1999})\BibitemShut {NoStop}%
\bibitem [{\citenamefont {Hamm}\ and\ \citenamefont {Zanni}(2011)}]{Hamm}%
  \BibitemOpen
  \bibfield  {author} {\bibinfo {author} {\bibfnamefont {P.}~\bibnamefont
  {Hamm}}\ and\ \bibinfo {author} {\bibfnamefont {M.}~\bibnamefont {Zanni}},\
  }\href {\doibase 10.1017/CBO9780511675935} {\emph {\bibinfo {title} {Concepts
  and Methods of 2D Infrared Spectroscopy}}}\ (\bibinfo  {publisher} {Cambridge
  University Press},\ \bibinfo {year} {2011})\BibitemShut {NoStop}%
\bibitem [{\citenamefont {Cho}(2009)}]{Cho}%
  \BibitemOpen
  \bibfield  {author} {\bibinfo {author} {\bibfnamefont {M.}~\bibnamefont
  {Cho}},\ }\href@noop {} {\emph {\bibinfo {title} {{Two-Dimensional Optical
  Spectroscopy}}}}\ (\bibinfo  {publisher} {CRC Press},\ \bibinfo {year}
  {2009})\BibitemShut {NoStop}%
\bibitem [{\citenamefont {Tanimura}\ and\ \citenamefont
  {Mukamel}(1993)}]{Inhomo01}%
  \BibitemOpen
  \bibfield  {author} {\bibinfo {author} {\bibfnamefont {Y.}~\bibnamefont
  {Tanimura}}\ and\ \bibinfo {author} {\bibfnamefont {S.}~\bibnamefont
  {Mukamel}},\ }\href {\doibase 10.1063/1.465484} {\bibfield  {journal}
  {\bibinfo  {journal} {Journal of Chemical Physics}\ }\textbf {\bibinfo
  {volume} {99}},\ \bibinfo {pages} {9496} (\bibinfo {year}
  {1993})}\BibitemShut {NoStop}%
\bibitem [{\citenamefont {Tanimura}\ and\ \citenamefont
  {Okumura}(1997)}]{AH01}%
  \BibitemOpen
  \bibfield  {author} {\bibinfo {author} {\bibfnamefont {Y.}~\bibnamefont
  {Tanimura}}\ and\ \bibinfo {author} {\bibfnamefont {K.}~\bibnamefont
  {Okumura}},\ }\href {\doibase 10.1063/1.473099} {\bibfield  {journal}
  {\bibinfo  {journal} {Journal of Chemical Physics}\ }\textbf {\bibinfo
  {volume} {106}},\ \bibinfo {pages} {2078} (\bibinfo {year}
  {1997})}\BibitemShut {NoStop}%
\bibitem [{\citenamefont {Okumura}\ and\ \citenamefont
  {Tanimura}(1997{\natexlab{a}})}]{AH02}%
  \BibitemOpen
  \bibfield  {author} {\bibinfo {author} {\bibfnamefont {K.}~\bibnamefont
  {Okumura}}\ and\ \bibinfo {author} {\bibfnamefont {Y.}~\bibnamefont
  {Tanimura}},\ }\href {\doibase 10.1063/1.474604} {\bibfield  {journal}
  {\bibinfo  {journal} {Journal of Chemical Physics}\ }\textbf {\bibinfo
  {volume} {107}},\ \bibinfo {pages} {2267} (\bibinfo {year}
  {1997}{\natexlab{a}})}\BibitemShut {NoStop}%
\bibitem [{\citenamefont {Okumura}\ and\ \citenamefont
  {Tanimura}(1997{\natexlab{b}})}]{AH03}%
  \BibitemOpen
  \bibfield  {author} {\bibinfo {author} {\bibfnamefont {K.}~\bibnamefont
  {Okumura}}\ and\ \bibinfo {author} {\bibfnamefont {Y.}~\bibnamefont
  {Tanimura}},\ }\href {\doibase 10.1016/S0009-2614(97)00832-4} {\bibfield
  {journal} {\bibinfo  {journal} {Chemical Physics Letters}\ }\textbf {\bibinfo
  {volume} {277}},\ \bibinfo {pages} {159} (\bibinfo {year}
  {1997}{\natexlab{b}})}\BibitemShut {NoStop}%
\bibitem [{\citenamefont {Tanimura}(1998)}]{AH04}%
  \BibitemOpen
  \bibfield  {author} {\bibinfo {author} {\bibfnamefont {Y.}~\bibnamefont
  {Tanimura}},\ }\href {\doibase 10.1016/S0301-0104(98)00010-X} {\bibfield
  {journal} {\bibinfo  {journal} {Chemical Physics}\ }\textbf {\bibinfo
  {volume} {233}},\ \bibinfo {pages} {217} (\bibinfo {year}
  {1998})}\BibitemShut {NoStop}%
\bibitem [{\citenamefont {Piryatinski}, \citenamefont {Chernyak},\ and\
  \citenamefont {Mukamel}(2001)}]{AH05}%
  \BibitemOpen
  \bibfield  {author} {\bibinfo {author} {\bibfnamefont {A.}~\bibnamefont
  {Piryatinski}}, \bibinfo {author} {\bibfnamefont {V.}~\bibnamefont
  {Chernyak}}, \ and\ \bibinfo {author} {\bibfnamefont {S.}~\bibnamefont
  {Mukamel}},\ }\href {\doibase 10.1016/S0301-0104(01)00253-1} {\bibfield
  {journal} {\bibinfo  {journal} {Chemical Physics}\ }\textbf {\bibinfo
  {volume} {266}},\ \bibinfo {pages} {311} (\bibinfo {year}
  {2001})}\BibitemShut {NoStop}%
\bibitem [{\citenamefont {Okumura}\ and\ \citenamefont
  {Tanimura}(1997{\natexlab{c}})}]{couple01}%
  \BibitemOpen
  \bibfield  {author} {\bibinfo {author} {\bibfnamefont {K.}~\bibnamefont
  {Okumura}}\ and\ \bibinfo {author} {\bibfnamefont {Y.}~\bibnamefont
  {Tanimura}},\ }\href {\doibase 10.1016/S0009-2614(97)00942-1} {\bibfield
  {journal} {\bibinfo  {journal} {Chemical Physics Letters}\ }\textbf {\bibinfo
  {volume} {278}},\ \bibinfo {pages} {175} (\bibinfo {year}
  {1997}{\natexlab{c}})}\BibitemShut {NoStop}%
\bibitem [{\citenamefont {Tokmakoff}\ \emph {et~al.}(1997)\citenamefont
  {Tokmakoff}, \citenamefont {Lang}, \citenamefont {Larsen}, \citenamefont
  {Fleming}, \citenamefont {Chernyak},\ and\ \citenamefont
  {Mukamel}}]{couple02}%
  \BibitemOpen
  \bibfield  {author} {\bibinfo {author} {\bibfnamefont {A.}~\bibnamefont
  {Tokmakoff}}, \bibinfo {author} {\bibfnamefont {M.~J.}\ \bibnamefont {Lang}},
  \bibinfo {author} {\bibfnamefont {D.~S.}\ \bibnamefont {Larsen}}, \bibinfo
  {author} {\bibfnamefont {G.~R.}\ \bibnamefont {Fleming}}, \bibinfo {author}
  {\bibfnamefont {V.}~\bibnamefont {Chernyak}}, \ and\ \bibinfo {author}
  {\bibfnamefont {S.}~\bibnamefont {Mukamel}},\ }\href {\doibase
  10.1103/PhysRevLett.79.2702} {\bibfield  {journal} {\bibinfo  {journal}
  {Physical Review Letters}\ }\textbf {\bibinfo {volume} {79}},\ \bibinfo
  {pages} {2702} (\bibinfo {year} {1997})}\BibitemShut {NoStop}%
\bibitem [{\citenamefont {Cho}, \citenamefont {Okumura},\ and\ \citenamefont
  {Tanimura}(1998)}]{couple03}%
  \BibitemOpen
  \bibfield  {author} {\bibinfo {author} {\bibfnamefont {M.}~\bibnamefont
  {Cho}}, \bibinfo {author} {\bibfnamefont {K.}~\bibnamefont {Okumura}}, \ and\
  \bibinfo {author} {\bibfnamefont {Y.}~\bibnamefont {Tanimura}},\ }\href
  {\doibase 10.1063/1.475505} {\bibfield  {journal} {\bibinfo  {journal}
  {Journal of Chemical Physics}\ }\textbf {\bibinfo {volume} {108}},\ \bibinfo
  {pages} {1326} (\bibinfo {year} {1998})}\BibitemShut {NoStop}%
\bibitem [{\citenamefont {Cho}(1999)}]{couple04}%
  \BibitemOpen
  \bibfield  {author} {\bibinfo {author} {\bibfnamefont {M.}~\bibnamefont
  {Cho}},\ }\href {\doibase 10.1063/1.479711} {\bibfield  {journal} {\bibinfo
  {journal} {Journal of Chemical Physics}\ }\textbf {\bibinfo {volume} {111}},\
  \bibinfo {pages} {4140} (\bibinfo {year} {1999})}\BibitemShut {NoStop}%
\bibitem [{\citenamefont {Okumura}, \citenamefont {Tokmakoff},\ and\
  \citenamefont {Tanimura}(1999)}]{couple05}%
  \BibitemOpen
  \bibfield  {author} {\bibinfo {author} {\bibfnamefont {K.}~\bibnamefont
  {Okumura}}, \bibinfo {author} {\bibfnamefont {A.}~\bibnamefont {Tokmakoff}},
  \ and\ \bibinfo {author} {\bibfnamefont {Y.}~\bibnamefont {Tanimura}},\
  }\href {\doibase 10.1063/1.479383} {\bibfield  {journal} {\bibinfo  {journal}
  {Journal of Chemical Physics}\ }\textbf {\bibinfo {volume} {111}},\ \bibinfo
  {pages} {492} (\bibinfo {year} {1999})}\BibitemShut {NoStop}%
\bibitem [{\citenamefont {Steffen}, \citenamefont {Fourkas},\ and\
  \citenamefont {Duppen}(1996)}]{dephase01}%
  \BibitemOpen
  \bibfield  {author} {\bibinfo {author} {\bibfnamefont {T.}~\bibnamefont
  {Steffen}}, \bibinfo {author} {\bibfnamefont {J.~T.}\ \bibnamefont
  {Fourkas}}, \ and\ \bibinfo {author} {\bibfnamefont {K.}~\bibnamefont
  {Duppen}},\ }\href {\doibase 10.1063/1.472594} {\bibfield  {journal}
  {\bibinfo  {journal} {Journal of Chemical Physics}\ }\textbf {\bibinfo
  {volume} {105}},\ \bibinfo {pages} {7364} (\bibinfo {year}
  {1996})}\BibitemShut {NoStop}%
\bibitem [{\citenamefont {Steffen}\ and\ \citenamefont
  {Duppen}(1998)}]{dephase02}%
  \BibitemOpen
  \bibfield  {author} {\bibinfo {author} {\bibfnamefont {T.}~\bibnamefont
  {Steffen}}\ and\ \bibinfo {author} {\bibfnamefont {K.}~\bibnamefont
  {Duppen}},\ }\href {\doibase 10.1016/S0301-0104(98)00083-4} {\bibfield
  {journal} {\bibinfo  {journal} {Chemical Physics}\ }\textbf {\bibinfo
  {volume} {233}},\ \bibinfo {pages} {267} (\bibinfo {year}
  {1998})}\BibitemShut {NoStop}%
\bibitem [{\citenamefont {Fourkas}(2007)}]{dephase03}%
  \BibitemOpen
  \bibfield  {author} {\bibinfo {author} {\bibfnamefont {J.~T.}\ \bibnamefont
  {Fourkas}},\ }\href {\doibase 10.1002/9780470141779.ch3} {\bibfield
  {journal} {\bibinfo  {journal} {Advances in Chemistry}\ }\textbf {\bibinfo
  {volume} {117}},\ \bibinfo {pages} {235} (\bibinfo {year}
  {2007})}\BibitemShut {NoStop}%
\bibitem [{\citenamefont {Okumura}\ and\ \citenamefont
  {Tanimura}(2003)}]{dephase04}%
  \BibitemOpen
  \bibfield  {author} {\bibinfo {author} {\bibfnamefont {K.}~\bibnamefont
  {Okumura}}\ and\ \bibinfo {author} {\bibfnamefont {Y.}~\bibnamefont
  {Tanimura}},\ }\href {\doibase 10.1021/jp027360o} {\bibfield  {journal}
  {\bibinfo  {journal} {Journal of Physical Chemistry A}\ }\textbf {\bibinfo
  {volume} {107}},\ \bibinfo {pages} {8092} (\bibinfo {year}
  {2003})}\BibitemShut {NoStop}%
\bibitem [{\citenamefont {Tanimura}\ and\ \citenamefont
  {Steffen}(2000{\natexlab{a}})}]{dephase05}%
  \BibitemOpen
  \bibfield  {author} {\bibinfo {author} {\bibfnamefont {Y.}~\bibnamefont
  {Tanimura}}\ and\ \bibinfo {author} {\bibfnamefont {T.}~\bibnamefont
  {Steffen}},\ }\href {\doibase 10.1143/JPSJ.69.4095} {\bibfield  {journal}
  {\bibinfo  {journal} {Journal of the Physical Society of Japan}\ }\textbf
  {\bibinfo {volume} {69}},\ \bibinfo {pages} {3115} (\bibinfo {year}
  {2000}{\natexlab{a}})}\BibitemShut {NoStop}%
\bibitem [{\citenamefont {Tanimura}\ and\ \citenamefont
  {Steffen}(2000{\natexlab{b}})}]{dephase06}%
  \BibitemOpen
  \bibfield  {author} {\bibinfo {author} {\bibfnamefont {Y.}~\bibnamefont
  {Tanimura}}\ and\ \bibinfo {author} {\bibfnamefont {T.}~\bibnamefont
  {Steffen}},\ }\href {\doibase 10.1143/JPSJ.69.4095} {\bibfield  {journal}
  {\bibinfo  {journal} {Journal of the Physical Society of Japan}\ }\textbf
  {\bibinfo {volume} {69}},\ \bibinfo {pages} {4095} (\bibinfo {year}
  {2000}{\natexlab{b}})}\BibitemShut {NoStop}%
\bibitem [{\citenamefont {Kato}\ and\ \citenamefont
  {Tanimura}(2002)}]{dephase07}%
  \BibitemOpen
  \bibfield  {author} {\bibinfo {author} {\bibfnamefont {T.}~\bibnamefont
  {Kato}}\ and\ \bibinfo {author} {\bibfnamefont {Y.}~\bibnamefont
  {Tanimura}},\ }\href {\doibase 10.1063/1.1503778} {\bibfield  {journal}
  {\bibinfo  {journal} {Journal of Chemical Physics}\ }\textbf {\bibinfo
  {volume} {117}},\ \bibinfo {pages} {6221} (\bibinfo {year}
  {2002})}\BibitemShut {NoStop}%
\bibitem [{\citenamefont {Kato}\ and\ \citenamefont
  {Tanimura}(2004)}]{dephase08}%
  \BibitemOpen
  \bibfield  {author} {\bibinfo {author} {\bibfnamefont {T.}~\bibnamefont
  {Kato}}\ and\ \bibinfo {author} {\bibfnamefont {Y.}~\bibnamefont
  {Tanimura}},\ }\href {\doibase 10.1063/1.1629272} {\bibfield  {journal}
  {\bibinfo  {journal} {Journal of Chemical Physics}\ }\textbf {\bibinfo
  {volume} {120}},\ \bibinfo {pages} {260} (\bibinfo {year}
  {2004})}\BibitemShut {NoStop}%
\bibitem [{\citenamefont {Tanimura}(2006)}]{dephase09}%
  \BibitemOpen
  \bibfield  {author} {\bibinfo {author} {\bibfnamefont {Y.}~\bibnamefont
  {Tanimura}},\ }\href {\doibase 10.1143/JPSJ.75.082001} {\bibfield  {journal}
  {\bibinfo  {journal} {Journal of the Physical Society of Japan}\ }\textbf
  {\bibinfo {volume} {75}},\ \bibinfo {pages} {82001} (\bibinfo {year}
  {2006})}\BibitemShut {NoStop}%
\bibitem [{\citenamefont {Blank}, \citenamefont {Kaufman},\ and\ \citenamefont
  {Fleming}(1999)}]{cascade01}%
  \BibitemOpen
  \bibfield  {author} {\bibinfo {author} {\bibfnamefont {D.~A.}\ \bibnamefont
  {Blank}}, \bibinfo {author} {\bibfnamefont {L.~J.}\ \bibnamefont {Kaufman}},
  \ and\ \bibinfo {author} {\bibfnamefont {G.~R.}\ \bibnamefont {Fleming}},\
  }\href {\doibase 10.1063/1.479591} {\bibfield  {journal} {\bibinfo  {journal}
  {Journal of Chemical Physics}\ }\textbf {\bibinfo {volume} {111}},\ \bibinfo
  {pages} {3105} (\bibinfo {year} {1999})}\BibitemShut {NoStop}%
\bibitem [{\citenamefont {Blank}, \citenamefont {Kaufman},\ and\ \citenamefont
  {Fleming}(2000)}]{cascade02}%
  \BibitemOpen
  \bibfield  {author} {\bibinfo {author} {\bibfnamefont {D.~A.}\ \bibnamefont
  {Blank}}, \bibinfo {author} {\bibfnamefont {L.~J.}\ \bibnamefont {Kaufman}},
  \ and\ \bibinfo {author} {\bibfnamefont {G.~R.}\ \bibnamefont {Fleming}},\
  }\href {\doibase 10.1063/1.481851} {\bibfield  {journal} {\bibinfo  {journal}
  {Journal of Chemical Physics}\ }\textbf {\bibinfo {volume} {113}},\ \bibinfo
  {pages} {771} (\bibinfo {year} {2000})}\BibitemShut {NoStop}%
\bibitem [{\citenamefont {Golonzka}\ \emph {et~al.}(2000)\citenamefont
  {Golonzka}, \citenamefont {Demird{\"{o}}ven}, \citenamefont {Khalil},\ and\
  \citenamefont {Tokmakoff}}]{cascade03}%
  \BibitemOpen
  \bibfield  {author} {\bibinfo {author} {\bibfnamefont {O.}~\bibnamefont
  {Golonzka}}, \bibinfo {author} {\bibfnamefont {N.}~\bibnamefont
  {Demird{\"{o}}ven}}, \bibinfo {author} {\bibfnamefont {M.}~\bibnamefont
  {Khalil}}, \ and\ \bibinfo {author} {\bibfnamefont {A.}~\bibnamefont
  {Tokmakoff}},\ }\href {\doibase 10.1063/1.1330236} {\bibfield  {journal}
  {\bibinfo  {journal} {Journal of Chemical Physics}\ }\textbf {\bibinfo
  {volume} {113}},\ \bibinfo {pages} {9893} (\bibinfo {year}
  {2000})}\BibitemShut {NoStop}%
\bibitem [{\citenamefont {Kaufman}\ \emph {et~al.}(2002)\citenamefont
  {Kaufman}, \citenamefont {Heo}, \citenamefont {Ziegler},\ and\ \citenamefont
  {Fleming}}]{CS2-01}%
  \BibitemOpen
  \bibfield  {author} {\bibinfo {author} {\bibfnamefont {L.~J.}\ \bibnamefont
  {Kaufman}}, \bibinfo {author} {\bibfnamefont {J.}~\bibnamefont {Heo}},
  \bibinfo {author} {\bibfnamefont {L.~D.}\ \bibnamefont {Ziegler}}, \ and\
  \bibinfo {author} {\bibfnamefont {G.~R.}\ \bibnamefont {Fleming}},\ }\href
  {\doibase 10.1103/PhysRevLett.88.207402} {\bibfield  {journal} {\bibinfo
  {journal} {Physical Review Letters}\ }\textbf {\bibinfo {volume} {88}},\
  \bibinfo {pages} {2074021} (\bibinfo {year} {2002})}\BibitemShut {NoStop}%
\bibitem [{\citenamefont {Kubarych}, \citenamefont {Milne},\ and\ \citenamefont
  {Miller}(2003{\natexlab{a}})}]{CS2-02}%
  \BibitemOpen
  \bibfield  {author} {\bibinfo {author} {\bibfnamefont {K.~J.}\ \bibnamefont
  {Kubarych}}, \bibinfo {author} {\bibfnamefont {C.~J.}\ \bibnamefont {Milne}},
  \ and\ \bibinfo {author} {\bibfnamefont {R.~J.}\ \bibnamefont {Miller}},\
  }\href {\doibase 10.1080/0144235031000121544} {\bibfield  {journal} {\bibinfo
   {journal} {International Reviews in Physical Chemistry}\ }\textbf {\bibinfo
  {volume} {22}},\ \bibinfo {pages} {497} (\bibinfo {year}
  {2003}{\natexlab{a}})}\BibitemShut {NoStop}%
\bibitem [{\citenamefont {Kubarych}\ \emph {et~al.}(2002)\citenamefont
  {Kubarych}, \citenamefont {Milne}, \citenamefont {Lin}, \citenamefont
  {Astinov},\ and\ \citenamefont {Miller}}]{CS2-03}%
  \BibitemOpen
  \bibfield  {author} {\bibinfo {author} {\bibfnamefont {K.~J.}\ \bibnamefont
  {Kubarych}}, \bibinfo {author} {\bibfnamefont {C.~J.}\ \bibnamefont {Milne}},
  \bibinfo {author} {\bibfnamefont {S.}~\bibnamefont {Lin}}, \bibinfo {author}
  {\bibfnamefont {V.}~\bibnamefont {Astinov}}, \ and\ \bibinfo {author}
  {\bibfnamefont {R.~J.}\ \bibnamefont {Miller}},\ }\href {\doibase
  10.1063/1.1429961} {\bibfield  {journal} {\bibinfo  {journal} {Journal of
  Chemical Physics}\ }\textbf {\bibinfo {volume} {116}},\ \bibinfo {pages}
  {2016} (\bibinfo {year} {2002})}\BibitemShut {NoStop}%
\bibitem [{\citenamefont {Kubarych}, \citenamefont {Milne},\ and\ \citenamefont
  {Miller}(2003{\natexlab{b}})}]{CS2-04}%
  \BibitemOpen
  \bibfield  {author} {\bibinfo {author} {\bibfnamefont {K.~J.}\ \bibnamefont
  {Kubarych}}, \bibinfo {author} {\bibfnamefont {C.~J.}\ \bibnamefont {Milne}},
  \ and\ \bibinfo {author} {\bibfnamefont {R.~J.}\ \bibnamefont {Miller}},\
  }\href {\doibase 10.1016/S0009-2614(03)00039-3} {\bibfield  {journal}
  {\bibinfo  {journal} {Chemical Physics Letters}\ }\textbf {\bibinfo {volume}
  {369}},\ \bibinfo {pages} {635} (\bibinfo {year}
  {2003}{\natexlab{b}})}\BibitemShut {NoStop}%
\bibitem [{\citenamefont {Milne}\ \emph {et~al.}(2006)\citenamefont {Milne},
  \citenamefont {Li}, \citenamefont {Jansen}, \citenamefont {Huang},\ and\
  \citenamefont {Miller}}]{C6H6-01}%
  \BibitemOpen
  \bibfield  {author} {\bibinfo {author} {\bibfnamefont {C.~J.}\ \bibnamefont
  {Milne}}, \bibinfo {author} {\bibfnamefont {Y.~L.}\ \bibnamefont {Li}},
  \bibinfo {author} {\bibfnamefont {T.~I.}\ \bibnamefont {Jansen}}, \bibinfo
  {author} {\bibfnamefont {L.}~\bibnamefont {Huang}}, \ and\ \bibinfo {author}
  {\bibfnamefont {R.~J.}\ \bibnamefont {Miller}},\ }\href {\doibase
  10.1021/jp062063v} {\bibfield  {journal} {\bibinfo  {journal} {Journal of
  Physical Chemistry B}\ }\textbf {\bibinfo {volume} {110}},\ \bibinfo {pages}
  {19867} (\bibinfo {year} {2006})}\BibitemShut {NoStop}%
\bibitem [{\citenamefont {Li}\ \emph {et~al.}(2008)\citenamefont {Li},
  \citenamefont {Huang}, \citenamefont {Miller}, \citenamefont {Hasegawa},\
  and\ \citenamefont {Tanimura}}]{HCONH2-01}%
  \BibitemOpen
  \bibfield  {author} {\bibinfo {author} {\bibfnamefont {Y.~L.}\ \bibnamefont
  {Li}}, \bibinfo {author} {\bibfnamefont {L.}~\bibnamefont {Huang}}, \bibinfo
  {author} {\bibfnamefont {R.~J.}\ \bibnamefont {Miller}}, \bibinfo {author}
  {\bibfnamefont {T.}~\bibnamefont {Hasegawa}}, \ and\ \bibinfo {author}
  {\bibfnamefont {Y.}~\bibnamefont {Tanimura}},\ }\href {\doibase
  10.1063/1.2927311} {\bibfield  {journal} {\bibinfo  {journal} {Journal of
  Chemical Physics}\ }\textbf {\bibinfo {volume} {128}},\ \bibinfo {pages}
  {234507} (\bibinfo {year} {2008})}\BibitemShut {NoStop}%
\bibitem [{\citenamefont {Hamm}\ and\ \citenamefont
  {Shalit}(2017)}]{2DTR-perspective}%
  \BibitemOpen
  \bibfield  {author} {\bibinfo {author} {\bibfnamefont {P.}~\bibnamefont
  {Hamm}}\ and\ \bibinfo {author} {\bibfnamefont {A.}~\bibnamefont {Shalit}},\
  }\href {\doibase 10.1063/1.4979288} {\bibfield  {journal} {\bibinfo
  {journal} {Journal of Chemical Physics}\ }\textbf {\bibinfo {volume} {146}},\
  \bibinfo {pages} {130901} (\bibinfo {year} {2017})}\BibitemShut {NoStop}%
\bibitem [{\citenamefont {Hamm}\ and\ \citenamefont {{J.
  Savolainen}}(2012)}]{2DTR-MD01}%
  \BibitemOpen
  \bibfield  {author} {\bibinfo {author} {\bibfnamefont {P.}~\bibnamefont
  {Hamm}}\ and\ \bibinfo {author} {\bibnamefont {{J. Savolainen}}},\ }\href
  {\doibase 10.1063/1.3691601} {\bibfield  {journal} {\bibinfo  {journal}
  {Journal of Chemical Physics}\ }\textbf {\bibinfo {volume} {136}},\ \bibinfo
  {pages} {094516} (\bibinfo {year} {2012})}\BibitemShut {NoStop}%
\bibitem [{\citenamefont {Hamm}\ \emph {et~al.}(2012)\citenamefont {Hamm},
  \citenamefont {Savolainen}, \citenamefont {Ono},\ and\ \citenamefont
  {Tanimura}}]{2DTR-MD02}%
  \BibitemOpen
  \bibfield  {author} {\bibinfo {author} {\bibfnamefont {P.}~\bibnamefont
  {Hamm}}, \bibinfo {author} {\bibfnamefont {J.}~\bibnamefont {Savolainen}},
  \bibinfo {author} {\bibfnamefont {J.}~\bibnamefont {Ono}}, \ and\ \bibinfo
  {author} {\bibfnamefont {Y.}~\bibnamefont {Tanimura}},\ }\href {\doibase
  10.1063/1.4729945} {\bibfield  {journal} {\bibinfo  {journal} {Journal of
  Chemical Physics}\ }\textbf {\bibinfo {volume} {136}},\ \bibinfo {pages}
  {236101} (\bibinfo {year} {2012})}\BibitemShut {NoStop}%
\bibitem [{\citenamefont {Hamm}(2014)}]{2DTR-MD03}%
  \BibitemOpen
  \bibfield  {author} {\bibinfo {author} {\bibfnamefont {P.}~\bibnamefont
  {Hamm}},\ }\href {\doibase 10.1063/1.4901216} {\bibfield  {journal} {\bibinfo
   {journal} {Journal of Chemical Physics}\ }\textbf {\bibinfo {volume}
  {141}},\ \bibinfo {pages} {184201} (\bibinfo {year} {2014})}\BibitemShut
  {NoStop}%
\bibitem [{\citenamefont {Savolainen}, \citenamefont {Ahmed},\ and\
  \citenamefont {Hamm}(2013)}]{2DTR-EX01}%
  \BibitemOpen
  \bibfield  {author} {\bibinfo {author} {\bibfnamefont {J.}~\bibnamefont
  {Savolainen}}, \bibinfo {author} {\bibfnamefont {S.}~\bibnamefont {Ahmed}}, \
  and\ \bibinfo {author} {\bibfnamefont {P.}~\bibnamefont {Hamm}},\ }\href
  {\doibase 10.1073/pnas.1317459110} {\bibfield  {journal} {\bibinfo  {journal}
  {Proceedings of the National Academy of Sciences}\ }\textbf {\bibinfo
  {volume} {110}},\ \bibinfo {pages} {20402} (\bibinfo {year}
  {2013})}\BibitemShut {NoStop}%
\bibitem [{\citenamefont {Ito}, \citenamefont {Hasegawa},\ and\ \citenamefont
  {Tanimura}(2014)}]{2DTR-MD04}%
  \BibitemOpen
  \bibfield  {author} {\bibinfo {author} {\bibfnamefont {H.}~\bibnamefont
  {Ito}}, \bibinfo {author} {\bibfnamefont {T.}~\bibnamefont {Hasegawa}}, \
  and\ \bibinfo {author} {\bibfnamefont {Y.}~\bibnamefont {Tanimura}},\ }\href
  {\doibase 10.1063/1.4895908} {\bibfield  {journal} {\bibinfo  {journal}
  {Journal of Chemical Physics}\ }\textbf {\bibinfo {volume} {141}},\ \bibinfo
  {pages} {124503} (\bibinfo {year} {2014})}\BibitemShut {NoStop}%
\bibitem [{\citenamefont {Pan}\ \emph {et~al.}(2015)\citenamefont {Pan},
  \citenamefont {Wu}, \citenamefont {Jin}, \citenamefont {Liu}, \citenamefont
  {Nagata}, \citenamefont {Zhang},\ and\ \citenamefont {Zhuang}}]{2DTR-MD05}%
  \BibitemOpen
  \bibfield  {author} {\bibinfo {author} {\bibfnamefont {Z.}~\bibnamefont
  {Pan}}, \bibinfo {author} {\bibfnamefont {T.}~\bibnamefont {Wu}}, \bibinfo
  {author} {\bibfnamefont {T.}~\bibnamefont {Jin}}, \bibinfo {author}
  {\bibfnamefont {Y.}~\bibnamefont {Liu}}, \bibinfo {author} {\bibfnamefont
  {Y.}~\bibnamefont {Nagata}}, \bibinfo {author} {\bibfnamefont
  {R.}~\bibnamefont {Zhang}}, \ and\ \bibinfo {author} {\bibfnamefont
  {W.}~\bibnamefont {Zhuang}},\ }\href {\doibase 10.1063/1.4917260} {\bibfield
  {journal} {\bibinfo  {journal} {Journal of Chemical Physics}\ }\textbf
  {\bibinfo {volume} {142}},\ \bibinfo {pages} {212419} (\bibinfo {year}
  {2015})}\BibitemShut {NoStop}%
\bibitem [{\citenamefont {Ikeda}, \citenamefont {Ito},\ and\ \citenamefont
  {Tanimura}(2015)}]{2DTR-TM01}%
  \BibitemOpen
  \bibfield  {author} {\bibinfo {author} {\bibfnamefont {T.}~\bibnamefont
  {Ikeda}}, \bibinfo {author} {\bibfnamefont {H.}~\bibnamefont {Ito}}, \ and\
  \bibinfo {author} {\bibfnamefont {Y.}~\bibnamefont {Tanimura}},\ }\href
  {\doibase 10.1063/1.4917033} {\bibfield  {journal} {\bibinfo  {journal}
  {Journal of Chemical Physics}\ }\textbf {\bibinfo {volume} {142}},\ \bibinfo
  {pages} {212421} (\bibinfo {year} {2015})}\BibitemShut {NoStop}%
\bibitem [{\citenamefont {Finneran}\ \emph {et~al.}(2016)\citenamefont
  {Finneran}, \citenamefont {Welsch}, \citenamefont {Allodi}, \citenamefont
  {Miller},\ and\ \citenamefont {Blake}}]{2DTR-EXA01}%
  \BibitemOpen
  \bibfield  {author} {\bibinfo {author} {\bibfnamefont {I.~A.}\ \bibnamefont
  {Finneran}}, \bibinfo {author} {\bibfnamefont {R.}~\bibnamefont {Welsch}},
  \bibinfo {author} {\bibfnamefont {M.~A.}\ \bibnamefont {Allodi}}, \bibinfo
  {author} {\bibfnamefont {T.~F.}\ \bibnamefont {Miller}}, \ and\ \bibinfo
  {author} {\bibfnamefont {G.~A.}\ \bibnamefont {Blake}},\ }\href {\doibase
  10.1073/pnas.1605631113} {\bibfield  {journal} {\bibinfo  {journal}
  {Proceedings of the National Academy of Sciences}\ }\textbf {\bibinfo
  {volume} {113}},\ \bibinfo {pages} {6857} (\bibinfo {year}
  {2016})}\BibitemShut {NoStop}%
\bibitem [{\citenamefont {Shalit}\ \emph {et~al.}(2017)\citenamefont {Shalit},
  \citenamefont {Ahmed}, \citenamefont {Savolainen},\ and\ \citenamefont
  {Hamm}}]{2DTR-EXC01}%
  \BibitemOpen
  \bibfield  {author} {\bibinfo {author} {\bibfnamefont {A.}~\bibnamefont
  {Shalit}}, \bibinfo {author} {\bibfnamefont {S.}~\bibnamefont {Ahmed}},
  \bibinfo {author} {\bibfnamefont {J.}~\bibnamefont {Savolainen}}, \ and\
  \bibinfo {author} {\bibfnamefont {P.}~\bibnamefont {Hamm}},\ }\href {\doibase
  10.1038/nchem.2642} {\bibfield  {journal} {\bibinfo  {journal} {Nature
  Chemistry}\ }\textbf {\bibinfo {volume} {9}},\ \bibinfo {pages} {273}
  (\bibinfo {year} {2017})}\BibitemShut {NoStop}%
\bibitem [{\citenamefont {Mead}\ \emph {et~al.}(2020)\citenamefont {Mead},
  \citenamefont {Lin}, \citenamefont {Magdau}, \citenamefont {{Miller III}},\
  and\ \citenamefont {Blake}}]{2DTR-EXD01}%
  \BibitemOpen
  \bibfield  {author} {\bibinfo {author} {\bibfnamefont {G.}~\bibnamefont
  {Mead}}, \bibinfo {author} {\bibfnamefont {H.-W.}\ \bibnamefont {Lin}},
  \bibinfo {author} {\bibfnamefont {I.-B.}\ \bibnamefont {Magdau}}, \bibinfo
  {author} {\bibfnamefont {T.~F.}\ \bibnamefont {{Miller III}}}, \ and\
  \bibinfo {author} {\bibfnamefont {G.~A.}\ \bibnamefont {Blake}},\ }\href
  {https://doi.org/10.1021/acs.jpcb.0c07935} {\bibfield  {journal} {\bibinfo
  {journal} {Journal of Physical Chemistry B}\ }\textbf {\bibinfo {volume}
  {124}},\ \bibinfo {pages} {8904} (\bibinfo {year} {2020})}\BibitemShut
  {NoStop}%
\bibitem [{\citenamefont {Ito}\ and\ \citenamefont {Tanimura}(2016)}]{2DIR}%
  \BibitemOpen
  \bibfield  {author} {\bibinfo {author} {\bibfnamefont {H.}~\bibnamefont
  {Ito}}\ and\ \bibinfo {author} {\bibfnamefont {Y.}~\bibnamefont {Tanimura}},\
  }\href {\doibase 10.1063/1.4941842} {\bibfield  {journal} {\bibinfo
  {journal} {Journal of Chemical Physics}\ }\textbf {\bibinfo {volume} {144}},\
  \bibinfo {pages} {74201} (\bibinfo {year} {2016})}\BibitemShut {NoStop}%
\bibitem [{\citenamefont {Grechko}\ \emph {et~al.}(2018)\citenamefont
  {Grechko}, \citenamefont {Hasegawa}, \citenamefont {D'Angelo}, \citenamefont
  {Ito}, \citenamefont {Turchinovich}, \citenamefont {Nagata},\ and\
  \citenamefont {Bonn}}]{2DIR-EX1}%
  \BibitemOpen
  \bibfield  {author} {\bibinfo {author} {\bibfnamefont {M.}~\bibnamefont
  {Grechko}}, \bibinfo {author} {\bibfnamefont {T.}~\bibnamefont {Hasegawa}},
  \bibinfo {author} {\bibfnamefont {F.}~\bibnamefont {D'Angelo}}, \bibinfo
  {author} {\bibfnamefont {H.}~\bibnamefont {Ito}}, \bibinfo {author}
  {\bibfnamefont {D.}~\bibnamefont {Turchinovich}}, \bibinfo {author}
  {\bibfnamefont {Y.}~\bibnamefont {Nagata}}, \ and\ \bibinfo {author}
  {\bibfnamefont {M.}~\bibnamefont {Bonn}},\ }\href {\doibase
  10.1038/s41467-018-03303-y} {\bibfield  {journal} {\bibinfo  {journal}
  {Nature Communications}\ }\textbf {\bibinfo {volume} {9}},\ \bibinfo {pages}
  {885} (\bibinfo {year} {2018})}\BibitemShut {NoStop}%
\bibitem [{\citenamefont {Nagata}\ and\ \citenamefont
  {Tanimura}(2006)}]{Nagata2DR1}%
  \BibitemOpen
  \bibfield  {author} {\bibinfo {author} {\bibfnamefont {Y.}~\bibnamefont
  {Nagata}}\ and\ \bibinfo {author} {\bibfnamefont {Y.}~\bibnamefont
  {Tanimura}},\ }\href {\doibase 10.1063/1.219185} {\bibfield  {journal}
  {\bibinfo  {journal} {Journal of Chemical Physics}\ }\textbf {\bibinfo
  {volume} {124}},\ \bibinfo {pages} {124508} (\bibinfo {year}
  {2006})}\BibitemShut {NoStop}%
\bibitem [{\citenamefont {Nagata}, \citenamefont {Hasegawa},\ and\
  \citenamefont {Tanimura}(2006)}]{Nagata2DR2}%
  \BibitemOpen
  \bibfield  {author} {\bibinfo {author} {\bibfnamefont {Y.}~\bibnamefont
  {Nagata}}, \bibinfo {author} {\bibfnamefont {T.}~\bibnamefont {Hasegawa}}, \
  and\ \bibinfo {author} {\bibfnamefont {Y.}~\bibnamefont {Tanimura}},\ }\href
  {\doibase 10.1063/1.2191850} {\bibfield  {journal} {\bibinfo  {journal}
  {Journal of Chemical Physics}\ }\textbf {\bibinfo {volume} {124}},\ \bibinfo
  {pages} {194504} (\bibinfo {year} {2006})}\BibitemShut {NoStop}%
\bibitem [{\citenamefont {Frostig}\ \emph {et~al.}(2015)\citenamefont
  {Frostig}, \citenamefont {Bayer}, \citenamefont {Dudovich}, \citenamefont
  {Eldar},\ and\ \citenamefont {Silberberg}}]{Frostig}%
  \BibitemOpen
  \bibfield  {author} {\bibinfo {author} {\bibfnamefont {H.}~\bibnamefont
  {Frostig}}, \bibinfo {author} {\bibfnamefont {T.}~\bibnamefont {Bayer}},
  \bibinfo {author} {\bibfnamefont {N.}~\bibnamefont {Dudovich}}, \bibinfo
  {author} {\bibfnamefont {Y.~C.}\ \bibnamefont {Eldar}}, \ and\ \bibinfo
  {author} {\bibfnamefont {Y.}~\bibnamefont {Silberberg}},\ }\href {\doibase
  10.1038/nphoton.2015.64} {\bibfield  {journal} {\bibinfo  {journal} {Nature
  Photonics}\ }\textbf {\bibinfo {volume} {9}},\ \bibinfo {pages} {339}
  (\bibinfo {year} {2015})}\BibitemShut {NoStop}%
\bibitem [{\citenamefont {Frostig}\ \emph {et~al.}(2017)\citenamefont
  {Frostig}, \citenamefont {Bayer}, \citenamefont {Eldar},\ and\ \citenamefont
  {Silberberg}}]{Frostig_signalrecovery}%
  \BibitemOpen
  \bibfield  {author} {\bibinfo {author} {\bibfnamefont {H.}~\bibnamefont
  {Frostig}}, \bibinfo {author} {\bibfnamefont {T.}~\bibnamefont {Bayer}},
  \bibinfo {author} {\bibfnamefont {Y.~C.}\ \bibnamefont {Eldar}}, \ and\
  \bibinfo {author} {\bibfnamefont {Y.}~\bibnamefont {Silberberg}},\ }\href
  {\doibase 10.1038/lsa.2017.115} {\bibfield  {journal} {\bibinfo  {journal}
  {Light: Science {\&} Applications}\ }\textbf {\bibinfo {volume} {6}},\
  \bibinfo {pages} {e17115} (\bibinfo {year} {2017})}\BibitemShut {NoStop}%
\bibitem [{\citenamefont {Silberberg}(2009)}]{Silberberg1}%
  \BibitemOpen
  \bibfield  {author} {\bibinfo {author} {\bibfnamefont {Y.}~\bibnamefont
  {Silberberg}},\ }\href {\doibase 10.1146/annurev.physchem.040808.090427}
  {\bibfield  {journal} {\bibinfo  {journal} {Annual Review of Physical
  Chemistry}\ }\textbf {\bibinfo {volume} {60}},\ \bibinfo {pages} {277}
  (\bibinfo {year} {2009})}\BibitemShut {NoStop}%
\bibitem [{\citenamefont {Evans}\ and\ \citenamefont {Xie}(2008)}]{tightfo}%
  \BibitemOpen
  \bibfield  {author} {\bibinfo {author} {\bibfnamefont {C.~L.}\ \bibnamefont
  {Evans}}\ and\ \bibinfo {author} {\bibfnamefont {X.~S.}\ \bibnamefont
  {Xie}},\ }\href {\doibase 10.1146/annurev.anchem.1.031207.112754} {\bibfield
  {journal} {\bibinfo  {journal} {Annual Review of Analytical Chemistry}\
  }\textbf {\bibinfo {volume} {1}},\ \bibinfo {pages} {883} (\bibinfo {year}
  {2008})}\BibitemShut {NoStop}%
\bibitem [{\citenamefont {Hurwitz}\ \emph {et~al.}(2020)\citenamefont
  {Hurwitz}, \citenamefont {Raanan}, \citenamefont {Ren}, \citenamefont
  {Frostig}, \citenamefont {Oulevey}, \citenamefont {Bruner}, \citenamefont
  {Dudovich},\ and\ \citenamefont {Silberberg}}]{Hurwitz}%
  \BibitemOpen
  \bibfield  {author} {\bibinfo {author} {\bibfnamefont {I.}~\bibnamefont
  {Hurwitz}}, \bibinfo {author} {\bibfnamefont {D.}~\bibnamefont {Raanan}},
  \bibinfo {author} {\bibfnamefont {L.}~\bibnamefont {Ren}}, \bibinfo {author}
  {\bibfnamefont {H.}~\bibnamefont {Frostig}}, \bibinfo {author} {\bibfnamefont
  {P.}~\bibnamefont {Oulevey}}, \bibinfo {author} {\bibfnamefont {B.~D.}\
  \bibnamefont {Bruner}}, \bibinfo {author} {\bibfnamefont {N.}~\bibnamefont
  {Dudovich}}, \ and\ \bibinfo {author} {\bibfnamefont {Y.}~\bibnamefont
  {Silberberg}},\ }\href {\doibase 10.1364/oe.384918} {\bibfield  {journal}
  {\bibinfo  {journal} {Optics Express}\ }\textbf {\bibinfo {volume} {28}},\
  \bibinfo {pages} {3803} (\bibinfo {year} {2020})}\BibitemShut {NoStop}%
\bibitem [{\citenamefont {Ito}, \citenamefont {Jo},\ and\ \citenamefont
  {Tanimura}(2015)}]{Hybrid02}%
  \BibitemOpen
  \bibfield  {author} {\bibinfo {author} {\bibfnamefont {H.}~\bibnamefont
  {Ito}}, \bibinfo {author} {\bibfnamefont {J.~Y.}\ \bibnamefont {Jo}}, \ and\
  \bibinfo {author} {\bibfnamefont {Y.}~\bibnamefont {Tanimura}},\ }\href
  {\doibase 10.1063/1.4932597} {\bibfield  {journal} {\bibinfo  {journal}
  {Structural Dynamics}\ }\textbf {\bibinfo {volume} {2}},\ \bibinfo {pages}
  {54102} (\bibinfo {year} {2015})}\BibitemShut {NoStop}%
\bibitem [{\citenamefont {Weiner}\ \emph {et~al.}(1990)\citenamefont {Weiner},
  \citenamefont {Leaird}, \citenamefont {Wiederrecht},\ and\ \citenamefont
  {Nelson}}]{Weiner1317}%
  \BibitemOpen
  \bibfield  {author} {\bibinfo {author} {\bibfnamefont {A.~M.}\ \bibnamefont
  {Weiner}}, \bibinfo {author} {\bibfnamefont {D.~E.}\ \bibnamefont {Leaird}},
  \bibinfo {author} {\bibfnamefont {G.~P.}\ \bibnamefont {Wiederrecht}}, \ and\
  \bibinfo {author} {\bibfnamefont {K.~A.}\ \bibnamefont {Nelson}},\ }\href
  {\doibase 10.1126/science.247.4948.1317} {\bibfield  {journal} {\bibinfo
  {journal} {Science}\ }\textbf {\bibinfo {volume} {247}},\ \bibinfo {pages}
  {1317} (\bibinfo {year} {1990})}\BibitemShut {NoStop}%
\bibitem [{\citenamefont {Weiner}(2011)}]{Weiner20113669}%
  \BibitemOpen
  \bibfield  {author} {\bibinfo {author} {\bibfnamefont {A.~M.}\ \bibnamefont
  {Weiner}},\ }\href {\doibase 10.1016/j.optcom.2011.03.084} {\bibfield
  {journal} {\bibinfo  {journal} {Optics Communications}\ }\textbf {\bibinfo
  {volume} {284}},\ \bibinfo {pages} {3669} (\bibinfo {year}
  {2011})}\BibitemShut {NoStop}%
\bibitem [{\citenamefont {Hasegawa}\ and\ \citenamefont
  {Tanimura}(2006)}]{Hybrid01}%
  \BibitemOpen
  \bibfield  {author} {\bibinfo {author} {\bibfnamefont {T.}~\bibnamefont
  {Hasegawa}}\ and\ \bibinfo {author} {\bibfnamefont {Y.}~\bibnamefont
  {Tanimura}},\ }\href {\doibase 10.1063/1.2217947} {\bibfield  {journal}
  {\bibinfo  {journal} {Journal of Chemical Physics}\ }\textbf {\bibinfo
  {volume} {125}},\ \bibinfo {pages} {74512} (\bibinfo {year}
  {2006})}\BibitemShut {NoStop}%
\bibitem [{\citenamefont {Chang}, \citenamefont {Peterson},\ and\ \citenamefont
  {Dang}(1995)}]{Chang:1995}%
  \BibitemOpen
  \bibfield  {author} {\bibinfo {author} {\bibfnamefont {T.~M.}\ \bibnamefont
  {Chang}}, \bibinfo {author} {\bibfnamefont {K.~A.}\ \bibnamefont {Peterson}},
  \ and\ \bibinfo {author} {\bibfnamefont {L.~X.}\ \bibnamefont {Dang}},\
  }\href {\doibase 10.1063/1.470319} {\bibfield  {journal} {\bibinfo  {journal}
  {Journal of Chemical Physics}\ }\textbf {\bibinfo {volume} {103}},\ \bibinfo
  {pages} {7502} (\bibinfo {year} {1995})}\BibitemShut {NoStop}%
\bibitem [{\citenamefont {Saito}\ and\ \citenamefont
  {Ohmine}(2003)}]{Saito:2003bp}%
  \BibitemOpen
  \bibfield  {author} {\bibinfo {author} {\bibfnamefont {S.}~\bibnamefont
  {Saito}}\ and\ \bibinfo {author} {\bibfnamefont {I.}~\bibnamefont {Ohmine}},\
  }\href {\doibase 10.1063/1.1609984} {\bibfield  {journal} {\bibinfo
  {journal} {Journal of Chemical Physics}\ }\textbf {\bibinfo {volume} {119}},\
  \bibinfo {pages} {9073} (\bibinfo {year} {2003})}\BibitemShut {NoStop}%
\bibitem [{\citenamefont {Olney}\ \emph {et~al.}(1997)\citenamefont {Olney},
  \citenamefont {Cann}, \citenamefont {Cooper},\ and\ \citenamefont
  {Brion}}]{Olney199759}%
  \BibitemOpen
  \bibfield  {author} {\bibinfo {author} {\bibfnamefont {T.~N.}\ \bibnamefont
  {Olney}}, \bibinfo {author} {\bibfnamefont {N.~M.}\ \bibnamefont {Cann}},
  \bibinfo {author} {\bibfnamefont {G.}~\bibnamefont {Cooper}}, \ and\ \bibinfo
  {author} {\bibfnamefont {C.~E.}\ \bibnamefont {Brion}},\ }\href {\doibase
  10.1016/S0301-0104(97)00145-6} {\bibfield  {journal} {\bibinfo  {journal}
  {Chemical Physics}\ }\textbf {\bibinfo {volume} {223}},\ \bibinfo {pages}
  {59} (\bibinfo {year} {1997})}\BibitemShut {NoStop}%
\bibitem [{\citenamefont {Hasegawa}\ and\ \citenamefont
  {Tanimura}(2011)}]{H-Water}%
  \BibitemOpen
  \bibfield  {author} {\bibinfo {author} {\bibfnamefont {T.}~\bibnamefont
  {Hasegawa}}\ and\ \bibinfo {author} {\bibfnamefont {Y.}~\bibnamefont
  {Tanimura}},\ }\href {\doibase 10.1021/jp111308f} {\bibfield  {journal}
  {\bibinfo  {journal} {Journal of Physical Chemistry B}\ }\textbf {\bibinfo
  {volume} {115}},\ \bibinfo {pages} {5545} (\bibinfo {year}
  {2011})}\BibitemShut {NoStop}%
\bibitem [{\citenamefont {Frostig}()}]{Hadasmaster}%
  \BibitemOpen
  \bibfield  {author} {\bibinfo {author} {\bibfnamefont {H.}~\bibnamefont
  {Frostig}},\ }\emph {\bibinfo {title} {{Single-pulse stimulated Raman
  scattering spectroscopy}}},\ \href {\doibase 10.1364/ol.36.001248} {Master's
  thesis},\ \bibinfo  {school} {Weizmann Institute of Science}\BibitemShut
  {NoStop}%
\bibitem [{\citenamefont {Jo}, \citenamefont {Ito},\ and\ \citenamefont
  {Tanimura}(2016)}]{JoCP2016}%
  \BibitemOpen
  \bibfield  {author} {\bibinfo {author} {\bibfnamefont {J.~Y.}\ \bibnamefont
  {Jo}}, \bibinfo {author} {\bibfnamefont {H.}~\bibnamefont {Ito}}, \ and\
  \bibinfo {author} {\bibfnamefont {Y.}~\bibnamefont {Tanimura}},\ }\href
  {\doibase 10.1016/j.chemphys.2016.07.002} {\bibfield  {journal} {\bibinfo
  {journal} {Chemical Physics}\ }\textbf {\bibinfo {volume} {481}},\ \bibinfo
  {pages} {245} (\bibinfo {year} {2016})}\BibitemShut {NoStop}%
\bibitem [{\citenamefont {Ito}, \citenamefont {Hasegawa},\ and\ \citenamefont
  {Tanimura}(2016)}]{IHTpol}%
  \BibitemOpen
  \bibfield  {author} {\bibinfo {author} {\bibfnamefont {H.}~\bibnamefont
  {Ito}}, \bibinfo {author} {\bibfnamefont {T.}~\bibnamefont {Hasegawa}}, \
  and\ \bibinfo {author} {\bibfnamefont {Y.}~\bibnamefont {Tanimura}},\ }\href
  {\doibase 10.1021/acs.jpclett.6b01766} {\bibfield  {journal} {\bibinfo
  {journal} {Journal of Physical Chemistry Letters}\ }\textbf {\bibinfo
  {volume} {7}},\ \bibinfo {pages} {4147} (\bibinfo {year} {2016})}\BibitemShut
  {NoStop}%
\bibitem [{\citenamefont {Liu}\ and\ \citenamefont {Liu}(2017)}]{JJLiu18}%
  \BibitemOpen
  \bibfield  {author} {\bibinfo {author} {\bibfnamefont {X.}~\bibnamefont
  {Liu}}\ and\ \bibinfo {author} {\bibfnamefont {J.}~\bibnamefont {Liu}},\
  }\href {https://www.tandfonline.com/doi/full/10.1080/00268976.2018.1434907}
  {\bibfield  {journal} {\bibinfo  {journal} {Molecular Physics}\ }\textbf
  {\bibinfo {volume} {116}},\ \bibinfo {pages} {755} (\bibinfo {year}
  {2017})}\BibitemShut {NoStop}%
\bibitem [{\citenamefont {Sakurai}\ and\ \citenamefont
  {Tanimura}(2011)}]{Sakurai-JPCA-2011-115}%
  \BibitemOpen
  \bibfield  {author} {\bibinfo {author} {\bibfnamefont {A.}~\bibnamefont
  {Sakurai}}\ and\ \bibinfo {author} {\bibfnamefont {Y.}~\bibnamefont
  {Tanimura}},\ }\href {\doibase 10.1021/jp1095618} {\bibfield  {journal}
  {\bibinfo  {journal} {Journal of Physical Chemistry A}\ }\textbf {\bibinfo
  {volume} {115}},\ \bibinfo {pages} {4009} (\bibinfo {year}
  {2011})}\BibitemShut {NoStop}%
\bibitem [{\citenamefont {Frostig}\ \emph {et~al.}(2011)\citenamefont
  {Frostig}, \citenamefont {Katz}, \citenamefont {Natan},\ and\ \citenamefont
  {Silberberg}}]{hadasmaster2}%
  \BibitemOpen
  \bibfield  {author} {\bibinfo {author} {\bibfnamefont {H.}~\bibnamefont
  {Frostig}}, \bibinfo {author} {\bibfnamefont {O.}~\bibnamefont {Katz}},
  \bibinfo {author} {\bibfnamefont {A.}~\bibnamefont {Natan}}, \ and\ \bibinfo
  {author} {\bibfnamefont {Y.}~\bibnamefont {Silberberg}},\ }\href {\doibase
  10.1364/ol.36.001248} {\bibfield  {journal} {\bibinfo  {journal} {Optics
  Letters}\ }\textbf {\bibinfo {volume} {36}},\ \bibinfo {pages} {1248}
  (\bibinfo {year} {2011})}\BibitemShut {NoStop}%
\bibitem [{\citenamefont {Rigneault}\ and\ \citenamefont {Berto}(2018)}]{CARS}%
  \BibitemOpen
  \bibfield  {author} {\bibinfo {author} {\bibfnamefont {H.}~\bibnamefont
  {Rigneault}}\ and\ \bibinfo {author} {\bibfnamefont {P.}~\bibnamefont
  {Berto}},\ }\href {\doibase 10.1063/1.5030335} {\bibfield  {journal}
  {\bibinfo  {journal} {APL Photonics}\ }\textbf {\bibinfo {volume} {3}},\
  \bibinfo {pages} {091101} (\bibinfo {year} {2018})}\BibitemShut {NoStop}%
\end{thebibliography}%

\end{document}